\documentclass[]{jfm}

\usepackage[pdftex,
           colorlinks=false,
           urlcolor=darkred,               
           filecolor=lightgrey,                 
           linkcolor=lightgrey,             
           citecolor=notsodarkblue,
           pdftitle={},
           pdfauthor={Dalton Harvie},
           pdfsubject={},
           pdfkeywords={},
           pdfborder={0 0 0},  
           pdfpagemode={UseThumbs},  
           bookmarks,
           bookmarksopen=true,
           bookmarksnumbered,
           letterpaper,
           raiselinks, 
           breaklinks,  
           linktocpage, 
           pdfstartview=FitH  
           ]{hyperref}
\usepackage{pdfcolmk}  

\definecolor{red}{rgb}{0.75,0,0}
\definecolor{green}{rgb}{0,0.75,0}
\definecolor{blue}{rgb}{0.24,0.12,0.74}
\definecolor{darkblue}{rgb}{0.09,0.05,0.29}
\definecolor{notsodarkblue}{rgb}{0.18,0.10,0.58}
\definecolor{darkred}{rgb}{0.54,0.,0.}
\definecolor{grey}{rgb}{0.32,0.32,0.32}
\definecolor{lightgrey}{rgb}{0.44,0.44,0.44}

\usepackage{amsmath}
\usepackage{amsfonts}
\usepackage{amssymb}

\usepackage[numbers]{natbib}

\usepackage{times}

\usepackage{framed}

\usepackage{subfig}

\usepackage{booktabs}

\usepackage{rotating}

\usepackage{supertabular}

\usepackage{floatpag}
\floatpagestyle{empty}
\rotfloatpagestyle{empty}

\newcommand{\scal}[2][]{#2_\mathrm{#1}}
\newcommand{\vect}[2][]{\boldsymbol{#2}_\mathrm{#1}}
\newcommand{\tens}[2][]{\mathsf{#2}_\mathrm{#1}}

\newcommand{\scali}[2][]{#2_{#1}}
\newcommand{\vecti}[2][]{\boldsymbol{#2}_{#1}}

\newcommand{\order}[1]{\ensuremath{\mathcal{O}\left(#1\right)}}
\newcommand{\cross}{\times}

\newcommand{\abs}[1]{\lvert#1\rvert}
\newcommand{\orders}[1]{\ensuremath{\mathcal{O}\left[#1\right]}}

\newcommand{\units}[1]{\ensuremath{\,\mathrm{#1} }}

\newlength{\explainwidth}
\newcommand{\explain}[2]{\settowidth{\explainwidth}{$#1$} \ensuremath{\underbrace{#1}_\text{\parbox{\explainwidth}{\centering \scriptsize #2}}}}

\newcommand{\circleme}[1]{\makebox[1.0em][l]{$\bigcirc$\hspace*{-0.7em}\scriptsize\textsf{#1}}}

\newcommand{\rightarea}{\overrightarrow{A}}
\newcommand{\rightsigma}{\overrightarrow{\sigma}}
\newcommand{\leftarea}{\overleftarrow{A}}
\newcommand{\leftsigma}{\overleftarrow{\sigma}}
\newcommand{\rightareaij}{\overrightarrow{A}_{ij}}
\newcommand{\leftareaij}{\overleftarrow{A}_{ij}}

\newcommand{\leftsigmasum}{$\underbrace{\rule[-1.0ex]{0ex}{0ex}\,\scal[12]{\sigma}\scal[12]{\leftarea}+\scal[1s]{\sigma}\scal[1s]{\leftarea}+\scal[2s]{\sigma}\scal[2s]{\leftarea}\,}_{\mbox{\rule[0ex]{0ex}{2.5ex}$\leftsigma\scal[cv]{A}$}}$}
\newcommand{\rightsigmasum}{$\underbrace{\rule[-1.0ex]{0ex}{0ex}\,\scal[1s]{\sigma}\scal[1s]{\rightarea}+\scal[2s]{\sigma}\scal[2s]{\rightarea}+\scal[12]{\sigma}\scal[12]{\rightarea}\,}_{\mbox{\rule[0ex]{0ex}{2.5ex}$\rightsigma\scal[cv]{A}$}}$}

\newcommand{\dst}{\widehat{t}}
\newcommand{\eqt}{\widetilde{t}}
\newcommand{\dsp}{\widehat{p}}
\newcommand{\eqp}{\widetilde{p}}
\newcommand{\dstk}{\dst_k}
\newcommand{\eqtk}{\eqt_k}
\newcommand{\dsDt}{\widehat{\Delta t}}
\newcommand{\eqDt}{\widetilde{\Delta t}}
\newcommand{\dsDtk}{{\dsDt}_k}
\newcommand{\eqDtk}{\eqDt_k}
\newcommand{\totT}{\overline{T}}
\newcommand{\eqT}{\widetilde{T}}
\newcommand{\dsT}{\widehat{T}}
\newcommand{\vcap}{\scal[cap]{v}}
\newcommand{\vcv}{\scal[cv]{v}}
\newcommand{\vvcv}{\vect[cv]{v}}
\newcommand{\vv}{\vect{v}}
\newcommand{\veqv}{\widetilde{\vect{v}}}
\newcommand{\vdsv}{\widehat{\vect{v}}}
\newcommand{\eqnablav}{\widetilde{\vnabla\vect{v}}}
\newcommand{\dsnablav}{\widehat{\vnabla\vect{v}}}
\newcommand{\rcv}{\scal[cv]{r}}
\newcommand{\lcv}{\scal[cv]{l}}
\newcommand{\Xcv}{\scal[cv]{X}}
\newcommand{\xcv}{\scal[cv]{x}}
\newcommand{\xcap}{\vect[cap]{x}}
\newcommand{\rcap}{\scal[cap]{r}}
\newcommand{\hrough}{\scal[rough]{h}}
\newcommand{\hroughcrit}{\scal[rough,crit]{h}}
\newcommand{\hsurround}{\scal[surround]{h}}
\newcommand{\hmol}{\scal[mol]{h}}
\newcommand{\taucap}{\scal[cap]{\tau}}
\newcommand{\thetaa}{\scal[a]{\theta}}
\newcommand{\thetar}{\scal[r]{\theta}}
\newcommand{\thetae}{\scal[e]{\theta}}
\newcommand{\Vcv}{\scal[cv]{V}}
\newcommand{\leftV}{\overleftarrow{V}}
\newcommand{\rightV}{\overrightarrow{V}}
\newcommand{\Acv}{\scal[cv]{A}}
\newcommand{\Scv}{\scal[cv]{S}}
\newcommand{\Scvcir}{\scal[cv,cir]{S}}
\newcommand{\Scvend}{\scal[cv,end]{S}}
\newcommand{\Scvbr}{\scal[cv,br]{S}}
\newcommand{\Scvbl}{\scal[cv,bl]{S}}
\newcommand{\Scvtop}{\scal[cv,top]{S}}
\newcommand{\Scvfluid}{\scal[cv,fluid]{S}}
\newcommand{\Vcvfluid}{\scal[cv,fluid]{V}}
\newcommand{\ncv}{\vect[cv]{n}}
\newcommand{\sigmaij}{\scali[ij]{\sigma}}
\newcommand{\Aij}{\scali[ij]{A}}
\newcommand{\nSij}{\vecti[\text{S,}ij]{n}}
\newcommand{\deltaSij}{\scali[\text{S,}ij]{\delta}}
\newcommand{\dsDeltasigma}{\widehat{\Delta \sigma}}
\newcommand{\vnabla}{\vect{\nabla}}
\newcommand{\rcvgrav}{\scal[cv,grav]{r}}
\newcommand{\rcvmax}{\scal[cv,max]{r}}
\newcommand{\vcvke}{\scal[cv,ke]{v}}
\newcommand{\vcvvis}{\scal[cv,vis]{v}}
\newcommand{\vcvcap}{\scal[cv,cap]{v}}
\newcommand{\vcvkep}{\scal[cv,ke']{v}}
\newcommand{\vcvvisp}{\scal[cv,vis']{v}}
\newcommand{\vcvcapp}{\scal[cv,cap']{v}}

\title{Contact Angle Hysteresis on Rough Surfaces Part I: Mechanical Energy Balance Framework}

\author{Dalton J.E. Harvie\aff{1}
  \corresp{\email{daltonh@unimelb.edu.au}}
 }

\affiliation{\aff{1}Department of Chemical Engineering, University of Melbourne, Parkville, VIC, 3010, Australia}

\begin{document}
\maketitle

\begin{abstract}
Using as a starting point conservation of momentum, a multiphase mechanical energy balance equation is derived that accounts for multiple material phases and interfaces present within a moving control volume.  This balance is applied to a control volume that is anchored to a three phase contact line as it advances over the surface of a rough and chemically homogeneous solid.  Using semi-quantitative models for the material behaviour occurring within the control volume, an order-of-magnitude analysis is performed to find what terms within the balance are significant, producing an equation that can be used to predict contact angle hysteresis from a knowledge of interface dynamics occurring around the three phase contact line.  In addition to this equation, the theory also answers several questions that have been discussed within the wetting literature:  Namely that (static) contact angle hysteresis is a function of conditions around the three phase contact line, as opposed to the surrounding flow system;  That contact angle hysteresis results from interface `jumps' that dissipate energy, rather than directly from contact line deformation;  That interfacial dynamics is required to predict contact angle hysteresis, but that these dynamics should be interpreted via energy conservation, and;  That dynamic contact angles depend on kinetic energy transport around the three phase contact line, as well as local energy dissipation.  The framework has been derived using assumptions of incompressible Newtonian fluids, reversible interface formation and zero-strain solids --- future work could relax these assumptions to make the theory more generally applicable.
\end{abstract}


\section{Introduction}

The ability of a liquid to `wet' a solid is described by the angle that the liquid makes with the solid when the interface is stationary --- the static contact angle.  For a liquid in contact with an ideal, smooth and chemically homogeneous surface, this angle $\thetae$ is unique and given by Young's Equation.  Real surfaces, which are often rough on a variety of length scales, are chemically heterogeneous and/or involve some type of irreversible work of adhesion, display a range of equilibrium contact angles: The maximum is the static advancing angle $\thetaa$, above will the interface will advance, and the minimum the static receding angle, $\thetar$, below which it will recede.  The difference between these two angles is defined as the range of contact angle hysteresis (CAH).  CAH, $\thetaa$ and $\thetar$ are critical wetting parameters that determine (for example) how easily drops can move over solid surfaces, under what conditions liquid films will smoothly coat surfaces or whether gas injection will aid particle floatation.  Technologies that depend on CAH angles include established processes such as industrial separation devices or the wetting behaviour of fabrics, through to more novel processes such as transparent self-cleaning surfaces for solar power generation or low-energy liquid fuel separation membranes \citep{cassie44,wu02,feng04,sun05,callies05,li13}.  For the design and optimisation of these processes general and validated wetting theories are needed that can predict the CAH range.  However, as highlighted via several recent works such theories are not yet available, with fundamental questions remaining about the nature of the wetting process\citep{eral13,jiang19,butt22}.  The purpose of this study is to derive an energy conservation framework that can be applied to predict CAH angles.

The early energy-based wetting theories of Wenzel and Cassie remain influential in interpreting wetting phenomena.  Considering the energy change that occurs as a liquid/gas interface advances a small distance over the surface of a rough solid, \citet{wenzel36} proposed that the apparent contact angle is related to the roughness $r$ of the surface, defined as the total to projected surface area ratio (see section \ref{sec:summary}).  In deriving this theory, \citeauthor{wenzel36} assumed that the liquid completely wets each surface undulation.  \Citet{cassie44} recognised that such `complete' wetting did not necessarily occur, and derived an expression for the apparent contact angle on a partially wet surface in terms of the wetted and non-wetted liquid areas per projected solid area, being $f_1$ and $f_2$, respectively (see section \ref{sec:summary}).  The Wenzel and \citeauthor{cassie44} equations are useful for understanding and interpreting experimental data, however, theories based solely on these concepts \citep[e.g.][]{bico02,patankar03} are not predictive as the proportion of solid surface wetted by a liquid is not known \emph{a priori}.  Also, these theories give only one static contact angle for a rough structured surface, rather than the CAH range that is observed experimentally.

Another series of works is based on the concept of energy minimisation of an entire drop sitting on a rough surface.  \citet{johnson64} computed the free energy of drops residing at the centre of concentric sinusoidal roughness rings, showing that the energy of the system oscillated as the drop volume increased and the interface advanced over each ring.  They interpreted the amplitude of these energy oscillations as energy barriers that must be overcome by macroscopic vibrational energy to allow interface movement, implying that as the height of surface roughness decreases, the range of CAH should also decrease.  This conclusion is contrary to experimental evidence however which shows that roughness-induced CAH strongly depends on surface topology (relative shape), rather than absolute roughness size, provided that gravitational and Laplace pressure effects can be neglected on the lengthscale of the roughness \citep{oner00,dorrer08,li16a,jiang19}.  Other studies have used similar static free-energy minimisation concepts to explain CAH for a variety of periodically shaped surfaces\citep{extrand02,brandon03,marmur06,marmur22}, however in general the results do not qualitatively agree with observation.  For example, \citet{brandon03} used minimal surface energy modelling to show that the apparent contact angle range for a drop on a doubly periodic undulating surface approached a single value (the Cassie angle) as the drop size to roughness ratio increased, again contradicting the above referenced experimental observations that show that the CAH range becomes quite constant at large droplet to roughness size ratios.  A related question also remains about these analyses:  Is it really necessary to consider the energy of the entire flow system (most commonly a droplet) to calculate the CAH range, or is it instead a property associated with the three phase contact line (TPCL) that can be applied to a wide variety of surrounding flow systems?  This is an unresolved question that has garnered conflicting opinions \citep{gao07a,mchale07,gao07b,nosonovsky07,panchagnula07,marmur22}.

Other studies conclude that CAH is substantially a property of the TPCL region\citep{nosonovsky07,panchagnula07} and that CAH has its origin in the energy dissipation that occurs around the TPCL as it advances over the rough solid.  Central to this concept is that the advancing and receding angles are defined as those measured while the interface is moving (albeit at a vanishingly slow velocity) rather than being determined solely by static thermodynamic states.  An influential study in this vein is \citet{joanny84} who proposed a model for CAH on a surface that contains a dilute number of `strong defects' as a model for surface contamination or dilutely distributed surface roughness.  The theory considered the `pinning' and subsequent `depinning' or `jumping' of the contact line as it advanced over a surface, assuming that during each interface jump surface potential energy is dissipated to heat.  \Citeauthor{joanny84} calculated this energy dissipation amount under ideal conditions and incorporated it into an equation for CAH relevant to dilute defect surfaces.  More recent works have experimentally observed the pinning/depinning behaviour of the fluid interface near the TPCL\citep{jiang19,priest09,forsberg10}.  Other studies have used a variety of energy conservation principles to extend \citeauthor{joanny84}'s work to periodic surfaces\citep{raj12,butt17,jiang19} or interpreted measurements of CAH in terms contact line energy dissipation and interfacial `jumping' dynamics\citep{priest07,priest13,dorrer08,song22}.  Despite these successes however questions remain about this conceptual model of CAH;  around what specific TPCL region should energy be conserved, how should the energies of real rough surfaces that may contain randomly shaped structures or micro bubbles/droplets be incorporated in the energy analysis, and why must the energy dissipated due to interface jumping dynamics be balanced by only interfacial (rather than material) stress movements?


On a slightly different track \Citet{joanny84}'s work also explored the influence of 'weak' surface defects on CAH, referring to smooth defects as those that cause the fluid interface near the solid to become distorted, but that do not result in the aforementioned `pinning' and `depinning' behaviour of the TPCL.  \Citeauthor{joanny84}'s conclusion was that isolated weak defects do not generally result in hysteresis, however other works have extended this analysis to conclude that distributions of weak surface heterogeneities can cause CAH\citep{pomeau85,robbins87,opik00}.  In related work \citet{cox83} examined how an interface changes as it moves over a gently undulating sinusoidal periodic rough surface, showing that when the interface moved in the direction of roughness periodicity TPCL `jumps' occurred (i.e., `strong' defects leading to CAH), but when advancing in other directions relative to the periodicity direction the interface moved continuously (i.e., `weak' defects producing no CAH).  It should be noted that these theoretical studies predict the possible shapes that a fluid interface can take when passing over arrangements of surface heterogeneities, and from these the range of CAH angles are inferred:  In general the link between CAH angles calculated via these interface topology methods and those calculated by energy conservation has not been established.


Reviewing this body of literature key questions about wetting behaviour remain.  Specifically there is conjecture regarding how energy conservation, fluid interface distortion and the dynamics of interface `jumping' can be combined to predict CAH.  Critically, there is no experimental consensus or fundamental analysis that shows how energy conservation should be applied to predict CAH --- works that are based on energy conservation around entire droplets in general do not predict experimentally observed CAH trends, while studies that are based on energy changes around a moving TPCL lack a rigorous energy conservation basis.  Further, while many studies do view contact line jumping as a source of CAH (`strong' defects), there is confusion over whether fluid interface distortion (`weak' defects) in isolation can produce CAH, and how studies that predict CAH angles via interface topology and dynamics can be mathematically related to CAH angles resulting from energy conservation.

This study addresses these questions.  Specifically we derive a rigorous energy conservation framework which when applied to the moving TPCL can be used to predict contact angle hysteresis (CAH).  We do this by first deriving a general multiphase mechanical energy balance (MMEB) equation that (section \ref{sec:mmeb}) is equivalent to the conventional single phase mechanical energy balance equation but includes terms associated with interfacial stresses acting between each of the material phases.  We then apply this MMEB equation to a cylindrical control volume that is anchored to a TPCL as it moves across a rough solid surface at a vanishingly slow speed (section \ref{sec:advance}).  By adopting order-or-magnitude models that describe how each of the materials behave within the control volume during the advance, we neglect terms that are insignificant to the calculation and derive a resulting energy conservation equation that predicts CAH given knowledge of the interfacial dynamics occurring within the TPCL.  A strength of the analysis is that mathematical constraints are derived that determine under what physical conditions the theory will be valid (section \ref{sec:discussion}), giving insight into past and future modelling theories and experimental wetting studies.


\section{Macroscopic Mechanical Energy Balance for Immiscible Multiphase Mixtures \label{sec:mmeb}}

\begin{figure}[htbp]
\centering
\resizebox{0.7\textwidth}{!}{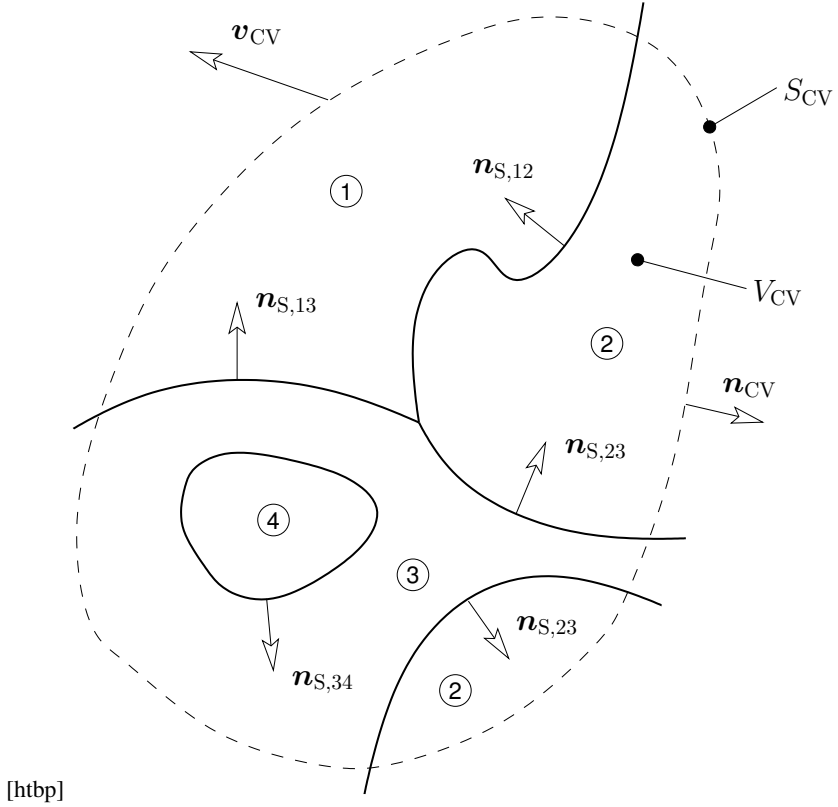} 
\caption{The macroscopic mechanical energy balance is performed on a moving control volume that contains multiple immiscible phases bounded by multiple interface types.  Each interface type has an associated unit normal vector $\vecti[\mathrm{S},ij]{n}$ that is directed into phase $i$ (where $i<j$).  The velocity and outward unit normal of the control volume boundary are $\vvcv$ and $\ncv$, respectively.  In the indicated configuration four material phases are present within the control volume, and they intersect along four different interface types.\label{fig:macroscopic_balance}}
\end{figure}

In this section we derive a macroscopic mechanical energy balance for a moving control volume that contains a number of immiscible phases (see figure \ref{fig:macroscopic_balance}).  The derivation is similar to that of \citet[][p221, \S 7.8]{bird02} except that interfacial tension acts at the interface between each pair of phases, and the balance is not specific to Newtonian liquids\footnote{We do assume that each material has a symmetric stress tensor, however for most homogeneous materials this assumption appears to be valid \citep{kuiken95,dahler61}.}.

Our starting point is a momentum equation for a mixture of immiscible phases,
\begin{equation}
\frac{\partial}{\partial t}\rho\vect{v} + \vnabla \cdot  \rho\vect{v}\vect{v} = \vnabla \cdot [ \tens[M]{T} + \tens[S]{T} ] + \rho\vect{g} \label{eq:momentum} .
\end{equation}
Here $\vect{v}$ is the local phase velocity (assumed to vary continuously throughout), $\rho$ the local phase density, $\tens[M]{T}$ the local total material stress at any point within any phase and $\tens[S]{T}$ the local surface stress acting on the interfaces between phases.  The functional form of the material stress tensor $\tens[M]{T}(\vect{x})$ is a property of the material type present at $\vect{x}$.

Formally, for equation (\ref{eq:momentum}) to be valid everywhere within $\Vcv$ all terms appearing in the equation must be defined not only within each phase but also on the interfaces between phases.  This includes terms such as $\tens[M]{T}$ and $\rho$ that are associated with a particular material type.  Such formal definitions could be made;  however as equation (\ref{eq:momentum}) is integrated over space in the following analysis, as long as any phase-specific terms are finite on each interface then their interface values do not affect the final energy balance.  Hence, we simply assume that the interface values for $\tens[M]{T}(\vect{x})$ and $\rho$ are finite.

The effects of interfacial tension on material movement are captured using a surface stress tensor, $\tens[S]{T}(\vect{x})$.  The form of this tensor is taken from \citet{lafaurie94}, but generalised here to include interfaces between multiple phases;
\begin{equation}
\tens[S]{T}(\vect{x}) = \sum_{i<j} \sigmaij ( \tens{I} - \vecti[\mathrm{S},ij]{n} \vecti[\mathrm{S},ij]{n} ) \scali[\mathrm{S},ij]{\delta} .
\label{eq:surface_stress}
\end{equation}
In this equation $\sigmaij$ is the constant surface energy per unit area (or interfacial tension) associated with the `$ij$ interface' (i.e., the interface between phases $i$ and $j$), $\tens{I}$ is the identity tensor, $\vecti[\mathrm{S},ij]{n}$ is a unit vector directed normal to the $ij$ interface and into phase $i$, and $\scali[\mathrm{S},ij]{\delta}$ is a `surface' delta function that is nonzero only on the $ij$ interface.  The surface delta function is essentially a multidimensional analogue of the Dirac delta function and has been utilised extensively in the development of computational fluid dynamics methods \citep{brackbill92,lafaurie94}.  In the present context it has the property that
\begin{equation}
\int_{\Vcv} \scali[\mathrm{S},ij]{\delta} dV = \scali[ij]{A}
\label{eq:area_definition} ,
\end{equation}
where $\scali[ij]{A}$ is the total area of the $ij$ interface existing within the control volume $\Vcv$.  Further, under conditions where $\ncv \cdot \vecti[\mathrm{S},ij]{n}$ is uniform over a particular intersection between a surface $\Scv$ and an interface defined by $\scali[\mathrm{S},ij]{\delta}$, the surface integral of the surface delta function is given by
\begin{equation}
\int_{\Scv} \scali[\mathrm{S},ij]{\delta} dS = \frac{\scali[ij]{l}}{\sqrt{1-(\ncv \cdot \vecti[\mathrm{S},ij]{n})^2}}
\label{eq:line_definition} ,
\end{equation}
where $\scali[ij]{l}$ represents the line length of the intersection between the two surfaces and $\ncv$ is a unit normal to the surface $\Scv$.  These properties of the surface delta function and others are discussed in more detail in the Appendix \ref{sec:deltaproperties}.

Note that in general interfacial stresses will act at each one of the immiscible material boundaries that exist within $\Vcv$.  Consequently, the sum in equation (\ref{eq:surface_stress}) cycles through all possible phase combinations under the condition that $i<j$; that is, $j=1\rightarrow m$ and $i=1\rightarrow j$ where $m$ is the total number of material phases present.  Thus, stresses from a possible $(m-1)!$ interface types may be included in the momentum balance.

By using equation (\ref{eq:surface_stress}) to represent surface stresses, three assumptions about the system are implied.  Firstly, as the surface stress is a sum of contributions from each interface type, we have neglected any `line tension' stresses that may occur at the intersection between interfaces.  While no concensus regarding the existence these stresses has been reached in the literature, most studies suggest that even if line tension does exist, it has a negligible effect on macroscopically measureable contact angles \citep{boruvka77,marmur97,pompe99,marmur06}.  Secondly, by assuming constant surface energies for each interface type, we have neglected any Marangoni forces that would exist if surfactants or thermal gradients were present within the control volume.  Thirdly, by representing the surface stress by equation (\ref{eq:surface_stress}) we have implicitly assumed that the process of surface creation or destruction is reversible on a molecular scale.  We discuss implications of this assumption in section \ref{sec:limitations}.

With the immiscible multiphase momentum equation defined, we proceed by taking the dot product of equation (\ref{eq:momentum}) with the local velocity $\vect{v}$ and then integrating the result over the volume $\Vcv$.  Noting that both the stress tensors $\tens[M]{T}$ and $\tens[S]{T}$ are symmetric, application of the Leibnitz formula for differentiating a volume integral, Gauss-Ostrogradskii theorem and compressible continuity equation yields,
\begin{multline}
\frac{d}{dt} \int_{\Vcv} \left ( \frac{1}{2} \rho v^2 + \rho \hat{\Phi} \right ) dV =
\int_{\Scv} \ncv \cdot \left [ \left ( \frac{1}{2} \rho v^2 + \rho \hat{\Phi} \right ) \left ( \vvcv - \vect{v} \right ) \right ] dS \\
+ \int_{\Scv} \ncv \cdot \left [ \tens[M]{T} \cdot \vect{v} \right ] dS - \int_{\Vcv} \tens[M]{T}:\vnabla \vect{v} dV \\
+ \int_{\Scv} \ncv \cdot \left [ \tens[S]{T} \cdot \vect{v} \right ] dS - \int_{\Vcv} \tens[S]{T}:\vnabla \vect{v} dV
\label{eq:conservation2} .
\end{multline}
Here $\vvcv$ and $\ncv$ are the velocity and outwardly directed unit normal of the control volume boundary $\Scv$, respectively, $v$ is the magnitude of the local velocity $\vect{v}$, and $\hat{\Phi}$ is a conservative gravitational potential function satisfying $\vect{g}=-\vnabla\hat{\Phi}$.

To simplify equation (\ref{eq:conservation2}) further we concentrate on the last two terms on the right hand side which relate to interfacial stresses.  For the first of these we substitute in the surface stress definition of equation (\ref{eq:surface_stress}) to find
\begin{equation}
\int_{\Scv} \ncv \cdot \left [ \tens[S]{T} \cdot \vect{v} \right ] dS = \sum_{i<j} \sigmaij \int_{\Scv}  \scali[\mathrm{S},ij]{\delta} \left ( \tens{I} - \vecti[\mathrm{S},ij]{n} \vecti[\mathrm{S},ij]{n} \right ):\vect{v} \ncv dS
\label{eq:surface_terms_1}
\end{equation}
For the second term we use the surface delta function transport equation derived in the Appendix (section \ref{sec:deltaproperties}),
\begin{equation}
\frac{\partial \scali[\mathrm{S},ij]{\delta}}{\partial t} + \vnabla \cdot \left ( \scali[\mathrm{S},ij]{\delta} \vect{v} \right ) = \scali[\mathrm{S},ij]{\delta} \left ( \tens{I} - \vecti[\mathrm{S},ij]{n} \vecti[\mathrm{S},ij]{n} \right ):\vnabla \vect{v}
\label{eq:delta_transport_equation} .
\end{equation}
Substituting $\tens[S]{T}$ from equation (\ref{eq:surface_stress}) into the second interfacial stress term of equation (\ref{eq:conservation2}), and then using the right hand side of the equation (\ref{eq:delta_transport_equation}) to expand the double dot product gives
\begin{align}
\int_{\Vcv} \tens[S]{T}:\vnabla \vect{v} dV & = \sum_{i<j} \int_{\Vcv} \sigmaij \scali[\mathrm{S},ij]{\delta} \left ( \tens{I} - \vecti[\mathrm{S},ij]{n} \vecti[\mathrm{S},ij]{n} \right ):\vnabla \vect{v} dV \nonumber \\
& = \sum_{i<j} \int_{\Vcv} \frac{\partial ( \sigmaij \scali[\mathrm{S},ij]{\delta} )}{\partial t} + \vnabla \cdot \left ( \sigmaij \scali[\mathrm{S},ij]{\delta} \vect{v} \right ) dV
\label{eq:surface_terms_2} .
\end{align}
Using the scalar Liebnitz theorem on the first term on the right of this equation and the Gauss-Ostrogradskii theorem on the second term yields
\begin{multline}
\int_{\Vcv} \tens[S]{T}:\vnabla \vect{v} dV = \\
\sum_{i<j} \left \{ \frac{d}{dt} \left ( \sigmaij \scali[ij]{A} \right ) - \int_{\Scv} \ncv \cdot \left [ \sigmaij \scali[\mathrm{S},ij]{\delta} ( \vvcv - \vect{v} ) \right ] dS \right \}
\label{eq:surface_terms_3}
\end{multline}
where equation (\ref{eq:area_definition}) has been used to relate surface area to the volume integral of $\scali[\mathrm{S},ij]{\delta}$.

Finally, substituting equations (\ref{eq:surface_terms_1}) and (\ref{eq:surface_terms_3}) back into equation (\ref{eq:conservation2}) and simplfying the material stress surface integral gives the immiscible multiphase mechanical energy balance valid for compressible and incompressible materials,
\begin{multline}
\explain{\frac{d}{dt} \left [ \int_{\Vcv} \left ( \frac{1}{2} \rho v^2 + \rho \hat{\Phi} \right ) dV + \sum_{i<j} \sigmaij \scali[ij]{A} \right ]}{rate of change of kinetic, gravitational potential and interfacial surface energy within $\Vcv$} = \\
%
\explain{\int_{\Scv} \ncv \cdot \left [ \left ( \frac{1}{2} \rho v^2 + \rho \hat{\Phi} + \sum_{i<j} \sigmaij \scali[\mathrm{S},ij]{\delta} \right ) \left ( \vvcv - \vect{v} \right ) \right ] dS}{rate at which kinetic, graviational potential and interfacial surface energy are advected into $\Vcv$} \\
%
+ \explain{\sum_{i<j} \sigmaij \int_{\Scv}  \scali[\mathrm{S},ij]{\delta} \left ( \tens{I} - \vecti[\mathrm{S},ij]{n} \vecti[\mathrm{S},ij]{n} \right ):\vect{v} \ncv dS }{rate of work done on the contents of $\Vcv$ by interfacial tension acting at $\Scv$}\\
+ \explain{\int_{\Scv} \tens[M]{T}:\vect{v} \ncv dS}{rate of work done on the contents of $\Vcv$ by material stresses acting at $\Scv$}%
- \explain{\int_{\Vcv} \tens[M]{T}:\vnabla \vect{v} dV}{rate at which energy dissipates to heat via material stresses acting within $\Vcv$}
\label{eq:macroscopic_balance}
\end{multline}
Along with the usual terms found in the single phase mechanical energy balance \citep[][p81, \S 3.3]{bird02} equation (\ref{eq:macroscopic_balance}) contains three interfacial stress terms:  The first represents the rate of change of interfacial energy contained within the control volume; the second the rate at which interfacial energy is advected across the control volume surface; and the third the rate at which interfacial stresses perform work on the control volume at the control volume surface.  Within the next section we demonstrate how these terms are evaluated for a specific control volume geometry.

\section{Calculating the advancing contact angle on a rough solid surface \label{sec:advance}}

In this section we analyse the macroscopic contact angle of an interface between two immiscible fluids that slowly advances over a rough solid.  The analysis uses the multiphase mechanical energy balance derived in section \ref{sec:mmeb}, applied to a small control volume (CV) which moves with the advancing three phase contact line across the surface of the solid.  Using semi-quantitative models for material and interface behaviour, an order of magnitude analysis is performed to determine which terms within the energy balance are significant, and from this an expression for the advancing contact angle is found that is valid in the limit of an infinitely slowly moving interface.

\subsection{Defining the physical system and moving control volume geometry \label{sec:physical_system}}

\begin{figure}[p]
\centering
\vspace*{-1.5cm}
\subfloat[\label{fig:three_phase_macro}]{\resizebox{0.5\textwidth}{!}{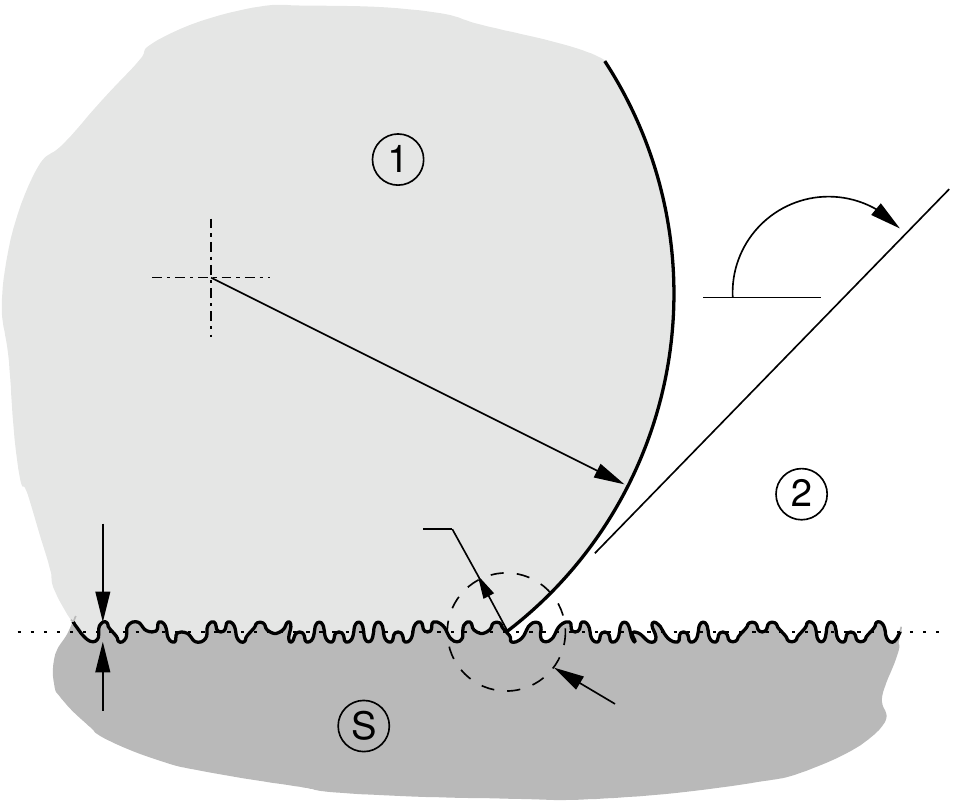}}
\vspace*{0.5cm}
\\
\subfloat[\label{fig:three_phase_micro}]{\resizebox{\textwidth}{!}{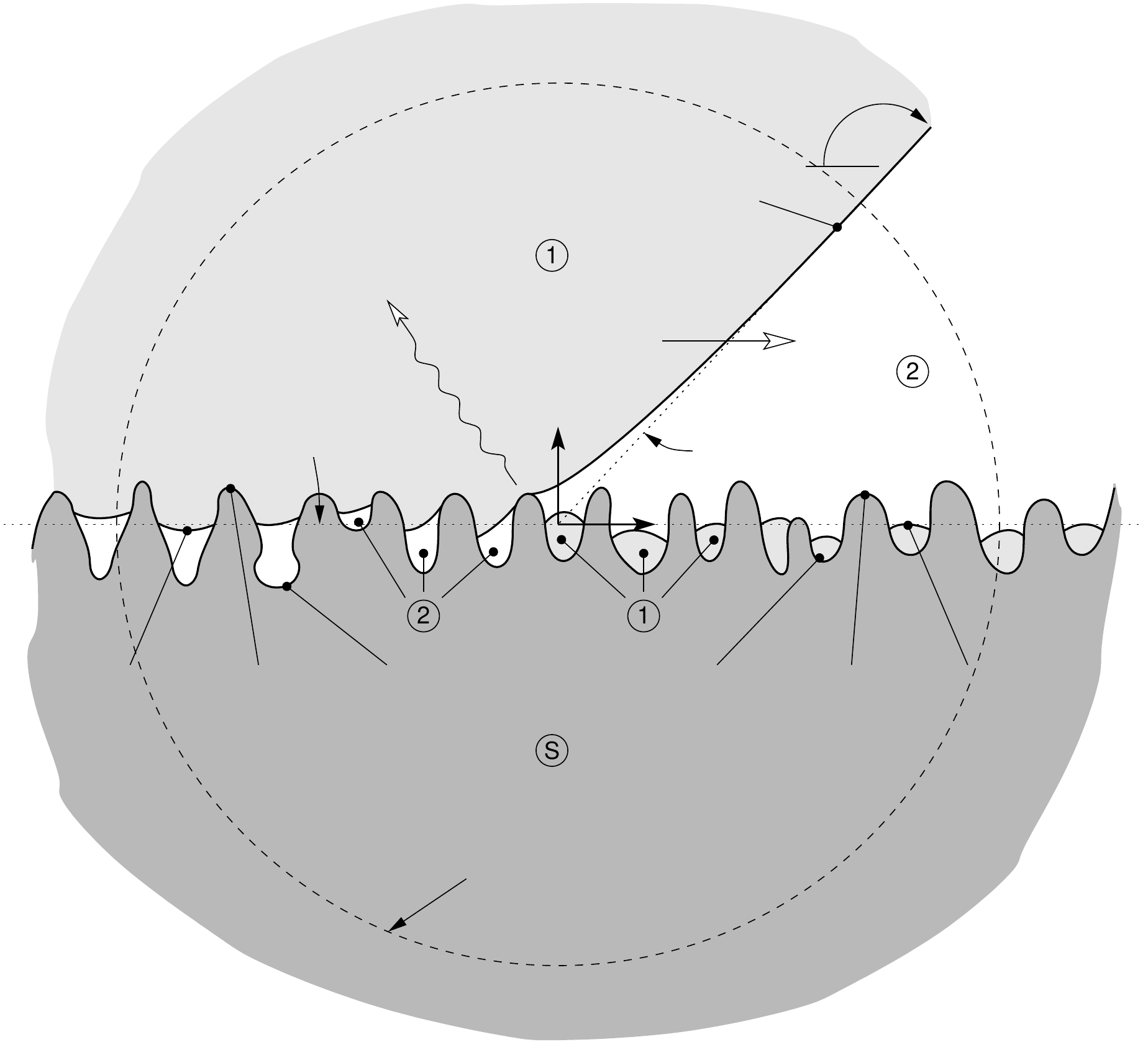}}
\caption{Frame (a) shows the region surrounding the control volume and contact line on a macroscopic scale, while frame (b) shows the same region on the scale of the control volume and solid surface roughness.  The symbols \protect\circleme{1}, \protect\circleme{2}, and \protect\circleme{S}, indicate regions of phase $1$ fluid, phase $2$ fluid, and solid, respectively.  \label{fig:three_phase}}
\end{figure}

Figure \ref{fig:three_phase} illustrates the physical system that is analysed.  Two fluid phases, labelled \circleme{1} and \circleme{2}, are bounded below by a solid phase, labelled \circleme{S}.  As the two fluids are immiscible they are separated by a distinct fluid interface.  The fluids may be either liquids or gases, but for this study are assumed to be completely insoluble with each other and with the solid, incompressible, and above a certain lengthscale $\hmol$ (for molecular), act as continua.  The implications of these assumptions are discussed further below and in section \ref{sec:discussion}.  The surface of the solid is rough, having undulations of a characteristic size $\hrough$.  A hydrodynamic flow is occuring on a lengthscale of $\hsurround$ which is much larger than $\hrough$.  This flow slowly drives the fluid interface to the right:  Hence, phase \circleme{1} is slowly advancing over the solid while phase \circleme{2} is slowly receding.

\begin{figure}
\centering
\resizebox{0.9\textwidth}{!}{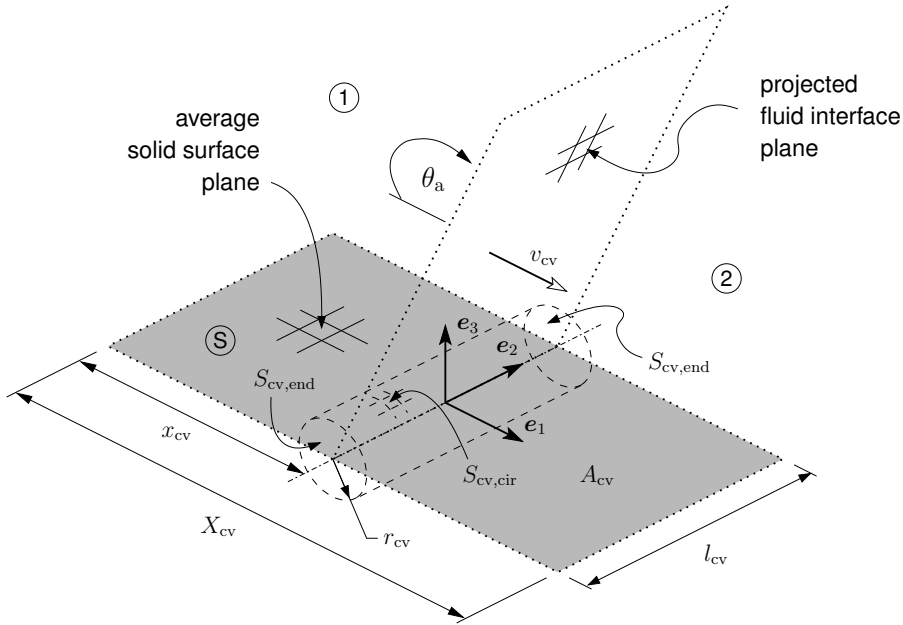}
\caption{The mechanical energy balance is performed over a cylindrical control volume that is located at the intersection of the average solid surface and projected fluid interface planes and moves forward over the solid at a speed of $\vcv$ and distance $\Xcv$.  \label{fig:three_phase_iso}}
\end{figure}

The multiphase mechanical energy balance is applied to a moving CV as it advances at a constant velocity $\vvcv=\vcv\vecti[1]{e}$ over a distance of $\Xcv$ along the solid surface.  This is illustrated in figure \ref{fig:three_phase_iso}.  Note that $\vcv$ is characteristic of the surrounding flow.  The CV contains and is approximately centred on the `three phase contact line' (TPCL), defined as the intersection between the advancing fluid interface and rough solid surface.  The moving CV has the geometry of a cylinder with radius $\rcv$ and length $\lcv$.  The dimensions of the volume are smaller than that of the surrounding hydrodynamic flow ($\hsurround$), yet larger than that of the solid surface roughness ($\hrough$).  Hence, noting the above description of the physical system and CV geometry we effectively assume the separation of four lengthscales in our analysis,
\begin{equation}
\hmol \ll \hrough \ll \rcv, \lcv, \Xcv \ll \hsurround .
\label{eq:lengthscales}
\end{equation}
Defining $\tau$ as the time taken for the CV to advance the entire distance $\Xcv$, it follows that $\Xcv=\vcv\tau$.

The precise centreline of the moving control volume is defined to lie at the intersection between two averaged planes:  the `average solid surface plane' and `projected fluid interface plane'.  The locations of these planes are defined as those of the solid surface and fluid interfaces, respectively, averaged over distances of $\order{\rcv}$ (where $\order{z}$ means `order $z$').  As the solid surface roughness $\hrough$ is of much smaller size than $\rcv$, it follows that the average solid surface plane is perfectly flat on the lengthscale of the CV.  In terms of the fluid interface, its topology is governed by the momentum and surface stress equations (equations (\ref{eq:momentum}) and (\ref{eq:surface_stress}), respectively), combined with boundary conditions specifying how the interface interacts with the solid surface.  The specific boundary condition that we employ for the microscopic contact angle is Youngs equation, expressed as
\begin{equation}
\left [ \vecti[{\text{S},12}]{n} \right ]_{\text{TPCL}} \cdot\vect[w]{n}=\cos\thetae
\label{eq:youngs_bc},
\end{equation}
where $\left [ \vecti[{\text{S},12}]{n}\right ]_{\text{TPCL}}$ is the unit normal to the fluid interface at a point on the three phase contact line, $\vect[w]{n}$ is the outwardly directed wall normal at the same contact point and $\thetae$ is the equilibrium or `Youngs' angle.  As equation (\ref{eq:youngs_bc}) specifies a direct relationship between the fluid interface ($\vecti[{\text{S},12}]{n}$) and solid ($\vect[w]{n}$) normals along the TPCL, it follows that close to the rough solid surface the fluid interface will have local curvatures that are characteristic of the solid roughness --- that is, of $\order{1/\hrough}$ --- and that these curvatures will exist within distances of $\order{\hrough}$ from the TPCL.  Conversely, further from the solid surface the topology of the interface varies over the larger lengthscales of the surrounding flow (indeed, this can be used to define $\hsurround$), so the curvature of the interface there approaches $\order{1/\hsurround}$.  Hence, as $\hrough \ll \rcv \ll \hsurround$ (equation (\ref{eq:lengthscales})), averaging the actual fluid interface over $\order{\rcv}$ produces a projected fluid interface plane that is perfectly flat on the lengthscales of the CV, and as the centreline of the control volume is defined as the intersection between the projected fluid interface plane and average solid surface plane, on the scale of the control volume its geometry is that of a perfect cylinder, with a perfectly straight centreline.

This description of the fluid interface topology and geometry of the CV has further implications for the mechanical energy balance application.  At distances of $\order{\rcv}$ from the TPCL, the actual fluid interface and projected fluid interface will at all times be coincident and perfectly flat on these lengthscales.  Hence, given that the projected fluid interface is used to define the centreline of the CV, where the actual fluid interface intersects the circumference of the CV (labeled as $\Scvcir$ in Figure \ref{fig:three_phase_iso}) it will be perfectly normal to the CV boundary at all times.  Similarly, along the ends of the CV (labelled as $\Scvend$ in Figure \ref{fig:three_phase_iso}) and at distances greater than $\order{\hrough}$ from the TPCL, the actual fluid interface will be flat, coincident with the projected fluid interface and perfectly normal to the CV boundary.  Closer to the TPCL however the interface will undulate with curvatures of $\order{1/\hrough}$, crossing the CV boundary at angles with are not necessarily normal to $\Scvend$.

The contact angle of phase \circleme{1} on the macroscopic lengthscale is defined as $\scal{\theta}$.  It is the angle between the average solid surface plane and fluid interface plane, measured through phase \circleme{1}.  Equivalently, consistent with the above, it is the angle between the average solid surface plane and actual fluid interface measured at distances of $\order{\rcv}$ from the TPCL.  The objective of our analysis is to determine the minimum macroscopic angle that \emph{just} causes the fluid interface to advance continually over the solid surface, albeit at the vanishingly slow velocity of $\vcv$.  An equivalent objective is to find the maximum macroscopic angle that \emph{just} allows the fluid interface to remain stationary.  Either definition represents the advancing angle of phase \circleme{1} over solid \circleme{S} in the presence of phase \circleme{2} and is referred to as $\thetaa$ (for `advancing').  Formally $\thetaa=\lim_{\vcv\rightarrow0} \theta$.  Note that as phase \circleme{1} advances over the solid, phase \circleme{2} recedes.  Thus an equivalent objective is to find the receding angle of phase \circleme{2} (equal to $\pi-\thetaa$).  Indeed, by swapping the physical properties between phases \circleme{1} and \circleme{2} (detailed in section \ref{sec:summary}), we can use the same analysis to determine the range of angles over which a fluid interface will remain stationary --- that is, the range of CAH.

\subsection{Describing material dynamics within the moving control volume \label{sec:material_dynamics}}

As well as defining the physical system and CV geometry used in the mechanical energy balance, to be able to perform an order of magnitude analysis on its various terms we need to quantitatively describe how the materials within the volume behave as a function of both space and time.  Specifically we need conceptual models for how the fluid velocities, pressures, interface topology and solid stresses vary as the CV advances.

\begin{figure}
\centering
\def\svgwidth{\textwidth}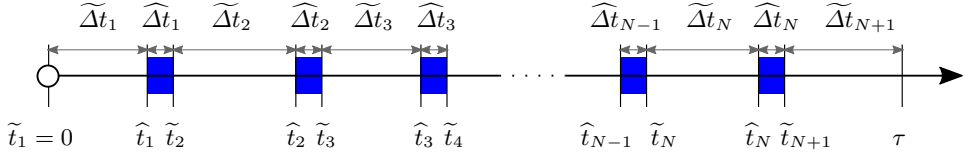
\caption{The entire analysis time $\tau$ is split into a number $N$ of dissipation periods (indicated in blue) where a portion of the TPCL moves at velocities of $\order{\vcap}$, interspersed between $N+1$ equilibrium stages where the entire TPCL moves at velocities of at most $\order{\vcv}$. \label{fig:timeline}}
\end{figure}

Within the fluid phase we assume that for the majority of the advancing time $\tau$ the TPCL and surrounding fluid both move continuously at the slow speeds of $\order{\vcv}$.  We refer to the system as being in `equilibrium' when this is the case and define the velocity field existing during these times as $\veqv=\order{\vcv}$.  However, at certain times during $\tau$, local areas of the TPCL will become pinned by particular surface defects, creating local interface deflections that become larger as the remainder of the TPCL continues to advance.  Eventually, once the surrounding TPCL has advanced some distance of $\order{\hrough}$ from the pinning location, these contact line sections will either detach from the surface defect and return to the main TPCL, or detach from the TPCL and form isolated bubbles/droplets of entrained fluid within the surface roughness.  Either way, these detachment processes cause the local TPCL and surrounding fluid to move at much faster capillary driven speeds than the continuous CV advance speed ($\vcv$), causing a viscous dissipation of energy.  This slip-stick dissipative motion has been described previously as the cause of contact angle hysteresis\citep{joanny84,raj12,butt17,jiang19} and as discussed in the Introduction has also been experimentally observed\citep{jiang19,priest09,forsberg10}.  In this study we define the local interface and fluid speeds associated with these capillary driven events as $\vdsv=\order{\vcap}$, and the total time during which there is a dissipation event occurring within the control volume as $\taucap$.  We further assume that the number of defects within the control volume ($N$) is small enough and the capillary velocity ($\vcap$) large enough that only one dissipation event occurs within the CV at any one time.  With these assumptions, and for convenience assuming that the analysis duration $\tau$ commences and finishes while the system is in equilibrium, we can split the total time over which the analysis is being conducted $\tau$ into a number of `dissipation events' ($N$) and `equilibrium stages' ($N+1$), with the $k$th dissipation event starting at $\dstk$ and lasting for $\dsDtk$, and the $k$th equilibrium stage starting at $\eqtk$ and lasting for $\eqDtk$.  The schematic timeline of Figure \ref{fig:timeline} illustrates this decomposition.  The following relationships result:
\begin{equation}
\begin{gathered}
\taucap=\sum_{k=1}^N \dsDtk \quad\quad \tau=\sum_{k=1}^{N+1} \eqDtk + \taucap \\
\dsDtk = \eqt_{k+1} - \dstk \quad\quad  \eqDtk = \dstk - \eqtk \\
\eqt_1=0 \quad\quad \eqt_{N+1} + \eqDt_{N+1} = \tau
\end{gathered}
\label{eq:time_decomposition}
\end{equation}
In terms of notation used in the remainder of the analysis, variables annotated with a `hat' correspond to properties associated with individual dissipation events (where parts of the TPCL are moving at $\order{\vcap}$), variables annotated with a `tilde' correspond to properties associated with the system while in equilibrium (where the entire TPCL is moving at at most $\order{\vcv}$), and variables annotated with a `bar' correspond to the entire advancing period $\tau$.  The decomposition of $\tau$ into separate dissipation and equilibrium stages is a key concept used in the subsequent energy analysis.

A number of mathematical constraints follow from the dynamic model of fluid movement described above.  Firstly, the analysis assumes that $\vcv \ll \vcap$.  By assuming that the fluids are Newtonian with a stress defined by
\begin{equation}
\tens[M]{T}=-p \tens{I} + \mu \left [ \vnabla \vect{v} + (\vnabla \vect{v})^T \right ]
\label{eq:newtonian_stress}
\end{equation}
an order of magnitude analysis on the augmented Navier-Stokes equations (as defined by equations (\ref{eq:momentum}) and (\ref{eq:surface_stress})) shows that capillary driven fluid velocities can be limited by either viscous dissipation or momentum acceleration/advection terms during any dissipation event.  Hence, a conservative estimate for the capillary driven velocity scale is given by equating the interfacial stress term to either the viscous dissipation or momentum acceleration/advection terms, giving
\begin{equation}
\vcap = \min \left ( \frac{\sigma}{\mu}, \sqrt{\frac{\sigma}{\rho\hrough}} \right )
\label{eq:vcap}
\end{equation}
In this expression and subsequent order of magnitude analyses, properties such as $\sigma$, $\mu$ and $\rho$ are order of magnitude estimates only, which for most expressions can be taken as the maximum of the different phase properties existing within the CV.  Equation (\ref{eq:vcap}) places a constraint on the maximum $\vcv$ that can be used given that $\vcv \ll \vcap$.

Interestingly, equation (\ref{eq:vcap}) predicts that capillary velocities are limited by momentum acceleration/advection terms rather than viscous dissipation on most practical surfaces.  To illustrate, for a water droplet within air advancing over a rough solid surface, using $\mu$ and $\rho$ from the water phase and $\sigma$ as the surface tension coefficient between air and water gives $\sigma/\mu \approx 72 \units{m/s}$.  However, for all surface roughness values $\hrough \gtrapprox \hroughcrit = \mu^2/(\rho \sigma) = 14 \units{nm}$ the capillary velocity will be limited by the momentum acceleration/advection term ($\sqrt{\sigma/(\rho\hrough)}$) and hence will determine $\vcap$ for this system.  Indeed, for a more typical surface roughness of $\hrough=10 \units{\mu m}$ equation (\ref{eq:vcap}) gives $\vcap \approx 3 \units{m/s}$ meaning that for practical surfaces as long as $\vcv$ is of the order cm/s or less $\vcv \ll \vcap$ will be satisfied.  We consider the variation of $\vcap$ in more detail in section \ref{sec:validity}.  Note that even though the local velocities existing during a dissipation event may be determined by a balance between capillary and momentum acceleration/advection terms, the Reynolds number for the motion near the surface roughness is not large ($\sim20$ for the above $\hrough=10 \units{\mu m}$ system).  This means that the size of the region where velocities are $\order{\vcap}$ during dissipation events is only of $\order{\hrough}$, and importantly these high velocities will not exist at the circumference of the CV, located at roughly $\rcv$ from the TPCL.

A second constraint required by the dynamic model of fluid movement outlined above is that $\taucap \ll \tau$.  To understand what conditions this places on the physical system we first recognise that the time taken for each dissipation event can be estimated from the interface velocity and distance travelled during each event --- $\dsDt_k=\order{\hrough/\vcap}$ --- and that as there are $N$ dissipation events occurring during $\tau$, $\taucap = \order{N\hrough/\vcap}$.  Recognising that the number of dissipation events occurring during the advance duration is $\order{\Acv/\hrough^2}$ where $\Acv = \Xcv \lcv$, and that $\tau=\Xcv/\vcv$ leads to
\begin{equation}
\frac{\taucap}{\tau}=\order{\frac{\lcv}{\hrough}\frac{\vcv}{\vcap}} \ll 1
\label{eq:taucapdtau}
\end{equation}
Equation (\ref{eq:taucapdtau}) can always be satisfied provided that $\vcv$ is small enough, which is obviously the case when determining $\thetaa$ as this angle is derived in the limit of $\vcv\rightarrow 0$.  If the energy balance is being applied to moving interfaces however (i.e., $\vcv \ne 0$) equation (\ref{eq:taucapdtau}) places a constraint on the maximum applicable $\vcv$.  This is discussed further in section \ref{sec:validity}.

Two final conceptual models concerning the continuous fluid movement occurring during the equilibrium stages of the flow have to be developed in order to apply the contact angle mechanical energy balance:  These models are for velocity gradient and pressure, both of which relate to the fluid stress.

For the velocity gradient, we note that during equilibrium stages the TPCL advances at a speed of $\order{\vcv}$ over the solid, resulting in a velocity discontinuity at the solid surface if the conventional continuum non-slip fluid boundary condition is applied.  Indeed, velocity profiles which satisfy the Navier-Stokes equations and that are consistent with both a moving TPCL and the non-slip boundary condition are available \citep{moffatt64,huh71}, however these result in velocity gradients near that TPCL that increase as $1/r$ (where $r$ is the distance to the TPCL).  We find that integrating these gradients over the region surrounding the TPCL in our energy balance results in a energy dissipation term for non-zero $\vcv$ that diverges logarithmically in an unphysical fashion, as others have found \citep{huh71}.  Solutions proposed to this problem, which we invoke here, all involve removing or limiting the stress (equivalently velocity gradients) within the fluid at small distances ($\hmol$) from the TPCL.  Various justifications for this limiting have been proposed \citep{huh71,joanny84,petrov92}, but most revolve around a breakdown of the Newtonian or continuum model of a fluid at the TPCL where individual molecules or particles within the fluid must `jump' along the solid.  The implications of this limit are discussed further in section \ref{sec:validity}.  Putting these concepts together we hence estimate the velocity gradients existing within the fluid during the equilibrium stages as
\begin{equation}
\eqnablav = \order{\frac{\vcv}{\max(r,\hmol)}}
\label{eq:gradv}
\end{equation}
where as discussed $\hmol$ is a small lengthscale related to the molecular (or non-continuum) nature of the fluid.

For the continuous flow pressure variation, we perform an order of magnitude on the single-phase Navier-Stokes equations, recognising that pressure gradients may develop within either fluid in response to both viscous stress and momentum acceleration/advection terms \footnote{Gravitational effects are not included when estimating the pressure variation as in general the fluid may not be at rest (and hence may not experience a hydrostatic pressure variation), however we note that inclusion of a gravitational contribution to the pressure variation would not alter the final results of the energy balance presented in section \ref{sec:finalcameb} as equivalent gravitational terms contribute to this balance via the gravitational potential energy transport term $\eqT_{3,k}$.}.  Given these assumptions we describe the pressure variation within either fluid phase as
\begin{align}
\widetilde{\vnabla p} & = \order{\rho \vnabla \cdot \veqv\veqv} + \order{\mu \vnabla^2 \veqv} \nonumber \\
& = \order{\frac{\rho \vcv^2}{\max(r,\hmol)}} + \order{\frac{\mu \vcv}{\left[ \max(r,\hmol) \right ]^2}}
\label{eq:pressure1}
\end{align}
where again the fluid stress has been limited within a distance of $\hmol$ from the TPCL.  Noting that the equilibrium stage fluid pressure $\eqp$ is relative to some point in the surrounding fluid far from the TPCL, and that under these slow flow conditions there is a potential pressure jump over the fluid interface due to the surrounding interface curvature of $\order{\sigma/\hsurround}$, we model the non-gravitational pressure variation within the CV during the equilibrium stages as
\begin{equation}
\eqp = \order{\rho \vcv^2} + \order{ \frac{\mu \vcv}{\max(r,\hmol)} } + \order{ \frac{\sigma}{\hsurround} } + p_0
\label{eq:pressure2}
\end{equation}
where $p_0$ is some reference pressure located at a point away from the TPCL, but within the vicinity of the CV.  All models describing the conceptual behaviour of the fluid phases during the equilibrium stages have now been defined.

For the solid phase, to apply the contact angle mechanical energy balance we need models that describe how the solid velocities and stresses vary within this phase as the TPCL advances over the rough surface.  In this study we invoke the simplest possible model by assuming that velocities within the solid are everywhere zero.  With this assumption the energy balance becomes independent of solid phase stresses.  On physical grounds zero velocities can be justified within the solid by assuming it is a yield-stress (or plastic) material that does not experience a stress exceeding its yield-stress during the analysis time.  In reality this is probably justifiable for most solids used in engineering applications, but for soft solids used in (eg) biomedical applications energy dissipation within the solid phase may be significant.  This is certainly an area for future work that could be incorporated into the presented mechanical energy balance framework but is not advanced here.

\subsection{Applying the mechanical energy balance to the moving control volume}

With the physical system defined and semi-quantitative models for how the materials within the CV behave as it advances over the rough solid, we can now apply the mechanical energy balance to find the advancing contact angle.

\subsubsection{Formulating the contact angle mechanical energy balance}

We start with deriving the most general form of the contact angle energy balance by applying equation (\ref{eq:macroscopic_balance}) to the moving CV and integrating it over a time period from $t_1$ to $t_2$, giving
\begin{equation}
T_0(t_1,t_2) = \sum_{i=1}^6 T_i(t_1,t_2)
\label{eq:balgen}
\end{equation}
where
\begin{align}
\scali[0]{T} (t_1,t_2) & = \frac{1}{\Acv} \left [ E(t=t_2)-E(t=t_1) \right ] \label{eq:T0} \\
E(t) & = \int_{\Vcv} \left ( \frac{1}{2} \rho v^2 + \rho \hat{\Phi} \right ) \, dV + \sum_{i<j} \sigmaij \Aij \label{eq:E} \\
\scali[1]{T} (t_1,t_2) & = \frac{1}{\Acv} \int^{t_2}_{t_1} \int_{\Scv} \sum_{i<j} \sigmaij \scali[\text{S},ij]{\delta} \ncv \cdot \vvcv \, dS dt \label{eq:T1} \\
\scali[2]{T} (t_1,t_2) & = \frac{1}{\Acv} \int^{t_2}_{t_1} \int_{\Scv} \ncv \cdot \frac{1}{2}\rho v^2 ( \vvcv - \vect{v} ) \, dS dt \label{eq:T2} \\
\scali[3]{T} (t_1,t_2) & = \frac{1}{\Acv} \int^{t_2}_{t_1} \int_{\Scv} \ncv \cdot \rho \hat{\Phi} ( \vvcv - \vect{v} ) \, dS dt \label{eq:T3} \\
\scali[4]{T} (t_1,t_2) & = - \frac{1}{\Acv} \int^{t_2}_{t_1} \int_{\Scv} \sum_{i<j} \sigmaij \deltaSij \nSij \nSij : \vect{v} \ncv \, dS dt \label{eq:T4} \\
\scali[5]{T} (t_1,t_2) & = \frac{1}{\Acv} \int^{t_2}_{t_1} \int_{\Scv} \tens[M]{T}:\vect{v} \ncv \, dS dt \label{eq:T5} \\
\scali[6]{T} (t_1,t_2) & = - \frac{1}{\Acv} \int^{t_2}_{t_1} \int_{\Vcv} \tens[M]{T}:\vnabla \vect{v} \, dV dt \label{eq:T6}
\end{align}
In the derivation use has been made of $\tens{I} : \vect{v}\ncv = \ncv \cdot \vect{v}$.

Next, defining the following notations that correspond to the entire, the $k$th equilibrium and the $k$th dissipation stages of the advance as
\begin{align}
\totT_i & = T_i(t_1=0,t_2=\tau) \label{eq:totT} \\
\eqT_{i,k} & = T_i(t_1=\eqtk,t_2=\eqtk+\eqDtk) \label{eq:eqT} \\
\dsT_{i,k} & = T_i(t_1=\dstk,t_2=\dstk+\dsDtk) \label{eq:dsT}
\end{align}
respectively for $i = 0$ to $6$, equation (\ref{eq:balgen}) is applied over the entire advance period from time $0$ to $\tau$ to give
\begin{equation}
\totT_0 = \sum_{i=1}^6 \totT_i
\label{eq:balbar}
\end{equation}
Recognising that $\totT_i$ for $i = 1$ to $6$ are all integrals over the total time period, using equation (\ref{eq:time_decomposition}) these terms can be written as sums of the corresponding terms from the equilibrium and dissipation stages, giving
\begin{equation}
\totT_0 = \sum_{i=1}^6 \left ( \sum_{k=1}^{N+1} \eqT_{i,k} + \sum_{k=1}^N \dsT_{i,k} \right ) = \sum_{i=1}^6 \sum_{k=1}^{N+1} \eqT_{i,k} + \sum_{k=1}^N \sum_{i=1}^6 \dsT_{i,k}
\label{eq:balbar2}
\end{equation}
To simplify the final term we apply the energy balance equation (\ref{eq:balgen}) to the $k$th dissipation period, giving
\begin{equation}
\dsT_{0,k} = \sum_{i=1}^6 \dsT_{i,k}
\label{eq:balhat2}
\end{equation}
which when substituted back into equation (\ref{eq:balbar2}) leads to
\begin{equation}
\totT_0 = \sum_{i=1}^6 \sum_{k=1}^{N+1} \eqT_{i,k} + \sum_{k=1}^N \dsT_{0,k}
\label{eq:balbar3}
\end{equation}
This is the form of the energy balance that is used to evaluate the advancing contact angle.  It expresses the total change in mechanical energy within the moving CV between the start and end of the advance as the sum of energy transfers happening during each of the equilibrium stages plus the change in mechanical energy existing within the CV that occurs over each of the dissipation events.  

We now examine each of the terms in equation (\ref{eq:balbar3}) finding either their order of magnitude, or for terms that prove to be significant, expressions that allow their evaluation in terms of system properties.

\subsubsection{Examining term $\totT_0$}

This term represents the change in mechanical energy within the CV between the start and end of the advance.  Defining the notation $\overline{\Delta a} = a(t=\tau) - a(t=0)$, $\totT_0$ can be expressed using equations (\ref{eq:T0}) and (\ref{eq:E}) as
\begin{align}
\totT_0 & = \frac{1}{\Acv} \overline{\Delta E}\nonumber \\
& = \frac{1}{\Acv} \int_{\Vcv} \frac{1}{2} \overline{\Delta(\rho v^2)} dV + \frac{1}{\Acv} \int_{\Vcv} \overline{\Delta\rho} \hat{\Phi} dV + \sum_{i<j} \frac{\sigmaij \overline{\Delta \Aij}}{\Acv} \label{eq:totT01}
\end{align}

The first term on the RHS (right hand side) of this equation captures changes to the kinetic energy within the CV between the start and end of the advance.  As both $t=0$ and $t=\tau$ are within equilibrium periods, changes to $\rho v^2$ within the CV between $t=0$ and $\tau$ will be limited to a fluid volume that is within $\order{\hrough}$ of the solid surface.  Further, as fluid velocities are $\order{\vcv}$ at both times, the first term can be evaluated as $\order{\rho\vcv^2 \hrough \rcv/\Xcv}$ after cancelling $\lcv$.  Similarly, for the second gravitational potential energy term, changes to $\rho$ are also limited to the same volume within $\order{\hrough}$ of the solid.  Recognising that the gravitational potential function satisfies $\hat{\Phi}=\vect{g} \cdot \vect{x}$, a maximum magnitude for this potential function within the CV is $\hat{\Phi}=\order{g \lcv}$ where $g=|\vect{g}|$ is the gravitational constant.  Finally, for the third term in equation (\ref{eq:totT01}), which represents the change in surface potential energy within the CV between the start and end of the advance, the change in area of each interface will be of $\order{ \hrough \lcv}$, being composed of changes to interfacial areas that occur around the TPCL, as well as changes to the average of each $\Aij$ associated with solid interfaces under each of the fluid phases due to the (possibly) random nature of the surface defects.

With these assumptions the change in mechanical energy over the duration of the advance is evaluated as
\begin{align}
\totT_0 & = \order{\frac{\rho\vcv^2 \rcv\hrough}{\Xcv}} + \order{\frac{\rho g \rcv \lcv \hrough}{\Xcv}} + \order{\frac{\sigma \hrough}{\Xcv}} \nonumber \\
& = \order{\rho\vcv^2 \hrough} + \order{\rho g \lcv \hrough} + \order{\frac{\sigma \hrough}{\Xcv}} \label{eq:totT02}
\end{align}
where in the last line we have used $\order{\rcv}=\order{\Xcv}$.  The above three terms are later compared to other terms present in equation (\ref{eq:balbar3}) to gauge their significance in determining the advancing contact angle $\thetaa$.

\subsubsection{Examining term $\sum_{k=1}^N \dsT_{0,k}$}

In a similar fashion to $\totT_0$, but here applied over each dissipation period, $\dsT_{0,k}$ represents the change in mechanical energy within the CV occurring over the period of the $k$th dissipation event.  Defining $\widehat{\Delta a}_k = a(t=\dstk+\dsDtk) - a(t=\dstk)$ this change in energy can be expressed as
\begin{align}
\dsT_{0,k} & = \frac{1}{\Acv} \widehat{\Delta E}_k \nonumber \\
& = \frac{1}{\Acv} \int_{\Vcv} \frac{1}{2} \widehat{\Delta(\rho v^2)}_k dV + \frac{1}{\Acv} \int_{\Vcv} \widehat{\Delta\rho}_k \hat{\Phi} dV + \sum_{i<j} \frac{\sigmaij \widehat{\Delta \Aij}_k}{\Acv} \label{eq:dsT01}
\end{align}
where $\widehat{\Delta \Aij}_k$ is the change in area of interface $ij$ that is contained within the CV and that occurs over the $k$th dissipation event.

In order to evaluate the magnitude of the terms appearing in equation (\ref{eq:dsT01}), we return to our conceptual model for how the fluid and interfaces behave during dissipation events.  During a dissipation event, an area of the TPCL `depins' from a particular surface defect and moves at a capillary-driven velocity to a new `equilibrium' interface position.  These dissipation events cause interfacial areas to change by $\order{\hrough^2}$, and as interfacial curvatures resulting from the roughness extend by $\order{\hrough}$ into the fluid, cause fluid properties to change within a volume of $\order{\hrough^3}$ near the TPCL.  Hence changes to the kinetic energy and density within a volume of $\order{\hrough^3}$ caused by each dissipation event will contribute to the first two terms on the RHS of equation (\ref{eq:dsT01}), while changes of $\order{\hrough^2}$ to the interfacial areas of each phase combination due to each dissipation event will contribute to the third term on the RHS of this equation.


Concurrently, over the duration of each dissipation event ($\dsDtk$) continuous movement of the TPCL still occurs across the rough surface, and this movement also contributes to the terms on the RHS of equation (\ref{eq:dsT01}).  Specifically, within $\order{\hrough}$ of the TPCL, or a volume of $\order{\dsDtk \vcv \lcv \hrough}$, there will be a change in kinetic energy and density of the fluid occurring due to the continuous TPCL movement which will add contributions to the first two terms on the RHS of equation (\ref{eq:dsT01}).  Similarly, for the third term on the RHS of equation (\ref{eq:dsT01}), there will also be a change in interfacial areas of $\order{\dsDtk \vcv \lcv}$ due to the continuous TPCL movement that also needs to be included.

Hence, summing changes due to both the specific dissipation event and continuous TPCL movement occurring during each dissipation period, the magnitude of $\dsT_{0,k}$ can be expressed as
\begin{align}
\dsT_{0,k} & = \order{\frac{\rho \vcv^2 \hrough^3}{\Xcv \lcv}} + \order{\frac{\rho \vcv^3 \hrough \dsDtk}{\Xcv}} \nonumber \\
& + \order{\frac{\rho g \hrough^4}{\Xcv \lcv}} + \order{\frac{\rho g \lcv \vcv \hrough \dsDtk}{\Xcv}} \nonumber \\
& + \order{\frac{\sigma \hrough^2}{\Xcv \lcv}} + \order{\frac{\sigma \vcv \dsDtk}{\Xcv}}
\label{eq:dsT02}
\end{align}
where each pair of the six terms corresponds to the first, second and third terms appearing on the RHS of equation (\ref{eq:dsT01}), respectively, with the first of each pair corresponding to changes caused by the specific capillary-driven dissipation event, and the second of each pair corresponding to the changes due to continuous TPCL movement that occurs during each dissipation event period.

Returning to equation (\ref{eq:balbar3}), it is actually the sum of $\dsT_{0,k}$ from all $N$ dissipation events that is required in the contact angle mechanical energy balance.  Performing this sum on equation (\ref{eq:dsT02}), while noting $\sum_{k=1}^N \dsDtk = \taucap$, $N=\order{\Xcv \lcv /\hrough^2}$ and $\vcv \tau = \Xcv$ leads to
\begin{align}
\sum_{k=1}^N \dsT_{0,k} & = \order{\rho \vcv^2 \hrough} + \order{\rho \vcv^2 \hrough \frac{\taucap}{\tau}} \nonumber \\
& + \order{\rho g \hrough^2} + \order{\rho g \lcv \hrough \frac{\taucap}{\tau}} \nonumber \\
& + \order{\sigma} + \order{\sigma \frac{\taucap}{\tau}}
\label{eq:dsT03}
\end{align}
Noting from equation (\ref{eq:taucapdtau}) that $\order{\taucap/\tau} \ll 1$, terms involving this ratio can be neglected in comparison to other terms, and recognising that the $\order{\sigma}$ term in the above originated from the final term of equation (\ref{eq:dsT01}), we arrive at
\begin{equation}
\sum_{k=1}^N \dsT_{0,k} = \order{\rho \vcv^2 \hrough} + \order{\rho g \hrough^2} + \dsDeltasigma
\label{eq:dsT04}
\end{equation}
Here the specific dissipation event surface energy change per area traversed has been defined as
\begin{equation}
\dsDeltasigma = \sum_{k=1}^N \sum_{i<j} \sigmaij \frac{\widehat{\Delta \Aij}_k}{\Acv}
\label{eq:dsDeltasigma}
\end{equation}
This variable represents the sum of changes to potential surface energies occurring within the CV due to all capillary-driven dissipation events, per projected area of solid traversed.  This variable has a magnitude of $\order{\sigma}$. It is similar to the $W_d$ term used in \citet{joanny84} and will become key in determining $\thetaa$ from the contact angle energy analysis.

The next six terms all correspond to energy transfers that occur during the equilibrium stages of the advance.

\subsubsection{Examining term $\sum_{k=1}^{N+1} \eqT_{1,k}$ \label{sec:eqT1}}

\begin{figure}
\centering
\def\svgwidth{0.8\textwidth}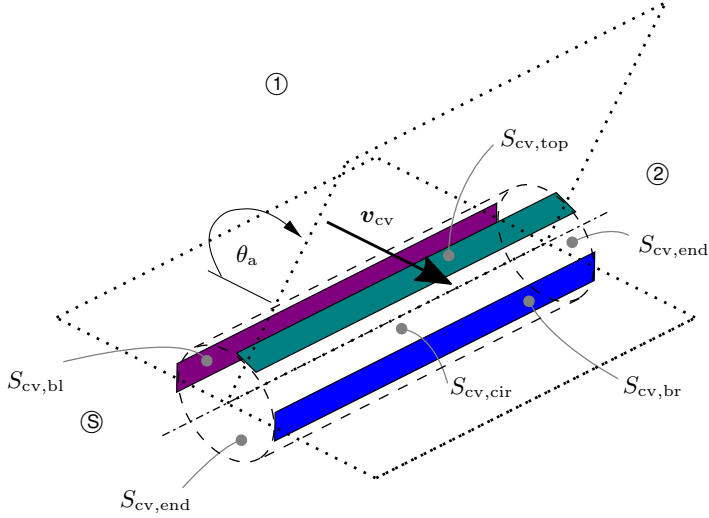
\caption{Three thin regions on the circumference of the CV ($\Scvcir$) are defined that contain all of the phase interfaces that intersect with this boundary. \label{fig:Scvcir_thinregions}}
\end{figure}

This term represents transport of surface potential energy through the boundary of the CV during the equilibrium stages due only to movement of the CV.  Introducing the shorthand notation $\int_{\eqDtk} = \int_{\eqtk}^{\eqtk+\eqDtk}$, applying equation (\ref{eq:T1}) over $\eqDtk$ gives
\begin{equation}
\sum_{k=1}^{N+1} \eqT_{1,k} = \frac{1}{\Acv} \sum_{k=1}^{N+1} \int_{\eqDtk} \int_{\Scv} \sum_{i<j} \sigmaij \scali[\text{S},ij]{\delta} \ncv \cdot \vvcv dS dt \label{eq:eqT1}
\end{equation}
Further, noting that $\ncv \cdot \vvcv$ on the ends of the CV ($\Scvend$), and that $\sigmaij$ is non-zero only within three thin regions on the circumference of the CV ($\Scvcir$) where the interfaces cross its boundary, equation (\ref{eq:eqT1}) can be written as the sum of three terms
\begin{align}
\sum_{k=1}^{N+1} \eqT_{1,k} & = \frac{1}{\Acv} \sum_{k=1}^{N+1} \int_{\eqDtk} \int_{\Scvtop} \scali[12]{\sigma} \scali[\text{S},12]{\delta} \ncv \cdot \vvcv dS dt \nonumber \\
& + \frac{1}{\Acv} \sum_{k=1}^{N+1} \int_{\eqDtk} \int_{\Scvbl} \sum_{i<j} \sigmaij \scali[\text{S},ij]{\delta} \ncv \cdot \vvcv dS dt \nonumber \\
& + \frac{1}{\Acv} \sum_{k=1}^{N+1} \int_{\eqDtk} \int_{\Scvbr} \sum_{i<j} \sigmaij \scali[\text{S},ij]{\delta} \ncv \cdot \vvcv dS dt 
\label{eq:eqT11}
\end{align}
Here, as indicated in Figure \ref{fig:Scvcir_thinregions}, $\Scvtop$, $\Scvbl$ and $\Scvbr$ are three thin regions located on $\Scvcir$ that just contain the fluid interface (top), interfaces associated with the solid surface under phase \circleme{1} (bl = bottom left) and interfaces associated with the solid surface under phase \circleme{2} (br = bottom right), respectively.

For the first term on the RHS of equation (\ref{eq:eqT11}), as $\ncv=-\cos \thetaa \vecti[1]{e} + \sin \thetaa \vecti[3]{e}$ where the fluid interface crosses the CV boundary at $\Scvtop$, and as $\vvcv=\vcv \vecti[1]{e}$, $\ncv \cdot \vvcv = -\vcv \cos \thetaa$ at this location.  Further, following the interface description outlined in Section \ref{sec:physical_system}, $\ncv \cdot \vecti[\mathrm{S},ij]{n} = 0$ at this location at all times, and as $\Scvtop$ is infinitely thin in the circumferential direction at this location equation (\ref{eq:line_definition}) shows that $\int_{\Scvtop} \deltaSij dS = \lcv$.  Hence the first term on the RHS of equation (\ref{eq:eqT11}) becomes
\begin{align}
\frac{1}{\Acv} \sum_{k=1}^{N+1} \int_{\eqDtk} \int_{\Scvtop} \scali[12]{\sigma} \deltaSij \ncv \cdot \vvcv dS dt & = -\frac{\scali[12]{\sigma} \vcv \cos \thetaa}{\Xcv} \sum_{k=1}^{N+1} \eqDtk \nonumber \\
& = - \left ( 1 - \frac{\taucap}{\tau} \right ) \scali[12]{\sigma} \cos \thetaa
\label{eq:eqT12}
\end{align}

The second and third terms on the RHS of equation (\ref{eq:eqT11}) represent the transport of surface potential energy associated with the rough solid surface out of, and into, respectively, the control volume as it advances across the solid.  Note that these surface energies may include contributions from micro-droplets or bubbles confined within the surface roughness, as well as from the surface energy between the solid and adjacent bulk fluid phase.

Focusing our explanation on the second term on the RHS of (\ref{eq:eqT11}) that involves an integral over $\Scvbl$, we first note that as $\Scvbl$ only has to include phase interfaces that are associated with the surface roughness, the circumferential height of $\Scvbl$ is only of $\order{\hrough}$. Further as $\hrough \ll \rcv$ (via equation (\ref{eq:lengthscales})) we have $\ncv = -\vecti[1]{e}$ at this location and consequently $\ncv \cdot \vvcv = -\vcv$.  Splitting the temporal integration using equation (\ref{eq:time_decomposition}) leads to
\begin{multline}
\frac{1}{\Acv} \sum_{k=1}^{N+1} \int_{\eqDtk} \int_{\Scvbl} \sum_{i<j} \sigmaij \deltaSij \ncv \cdot \vvcv dS dt = \\
-\frac{\vcv}{\Xcv \lcv} \sum_{i<j} \sigmaij \int_0^{\tau} \int_{\Scvbl} \deltaSij dS dt \\
- \frac{\vcv}{\Xcv \lcv} \sum_{i<j} \sigmaij \sum_{k=1}^{N+1} \int_{\dsDtk} \int_{\Scvbl} \scali[\text{S},ij]{\delta} dS dt
\label{eq:eqT13}
\end{multline}

For the first term on the RHS of this equation, we define a new volume $\leftV$ that is created by sweeping $\Scvbl$ over the solid surface for the duration of the advance.  Formally we define a coordinate $\xcv= t \vcv$ that increases in the direction of $\vecti[1]{e}$ and measures the progress of the CV as it moves over the solid surface, giving
\begin{align}
\vcv \sum_{i<j} \sigmaij \int_0^{\tau} \int_{\Scvbl} \deltaSij dS dt & = \sum_{i<j} \sigmaij \int_0^{\Xcv} \int_{\Scvbl} \deltaSij dS d\xcv \nonumber \\
& = \sum_{i<j} \sigmaij \int_{\leftV} \deltaSij dV \nonumber \\
& = \sum_{i<j} \sigmaij \leftareaij
\label{eq:eqT14}
\end{align}
where $\leftareaij$ is the area of each interface type $ij$ associated with the rough solid surface that leaves the CV during its advance over the solid.

For the second term on the RHS of equation (\ref{eq:eqT13}) we calculate its order of magnitude rather than derive an expression, noting that $\sum_{k=1}^N \dsDtk = \taucap$ and $\int_{\Scvbl} \scali[\text{S},ij]{\delta} dS = \order{\lcv}$.  Incorporating these expressions and equation (\ref{eq:eqT14}) back into equation (\ref{eq:eqT13}) gives
\begin{equation}
\frac{1}{\Acv} \sum_{k=1}^{N+1} \int_{\eqDtk} \int_{\Scvbl} \sum_{i<j} \sigmaij \deltaSij \ncv \cdot \vvcv dS dt =
- \sum_{i<j} \sigmaij \frac{\leftareaij}{\Acv} + \order{\frac{\taucap}{\tau}\sigma}
\label{eq:eqT15}
\end{equation}
completing the evaluation of the second term on the RHS of equation (\ref{eq:eqT11}).

Finally, performing an analogous calculation for the third term on the RHS of equation (\ref{eq:eqT11}) for the solid surface interfaces $\rightareaij$ that lie under phase \circleme{2} and that enter the CV through $\Scvbr$ during the advance, and noting again as per equation (\ref{eq:taucapdtau}) that terms involving the factor $\taucap/\tau$ can be neglected relative to those that do not include this term, the surface potential energy transport term from equation (\ref{eq:eqT11}) can be written as
\begin{equation}
\sum_{k=1}^{N+1} \eqT_{1,k} = - \scali[12]{\sigma} \cos \thetaa + \rightsigma - \leftsigma
\label{eq:eqT16}
\end{equation}
where the compound surface energies associated with the solid interface that is leaving (behind and to the left of the TPCL) and entering (in front and to the right of the TPCL) the advancing CV are defined by
\begin{equation}
\leftsigma = \sum_{i<j} \sigmaij \frac{\leftareaij}{\Acv} \quad \text{and} \quad \rightsigma = \sum_{i<j} \sigmaij \frac{\rightareaij}{\Acv} ,
\label{eq:areasigmas}
\end{equation}
respectively.  We note that these compound surface energies incorporate any micro-bubbles or droplets that could possibly be contained within the solid surface roughness.  For the leaving energy ($\leftsigma$) these could be formed during the discussed dissipation events that occur as the TPCL sweeps over rough solid.  For the entering energy ($\rightsigma$) these could be pre-existing within the surface roughness, possibly a result of previous wetting processes (such as receding back over a previously wetted surface).

\subsubsection{Examining term $\sum_{k=1}^{N+1} \eqT_{2,k}$}

This term represents the transport of kinetic energy into the CV during the equilibrium stages, and is given by
\begin{equation}
\sum_{k=1}^{N+1} \eqT_{2,k} = \frac{1}{\Acv} \sum_{k=1}^{N+1} \int_{\eqDtk} \int_{\Scv} \ncv \cdot \frac{1}{2}\rho v^2 ( \vvcv - \vv ) dS dt \label{eq:eqT21}
\end{equation}
Noting from section \ref{sec:material_dynamics} that velocities are zero within the solid phase and within the equilibrium stages within the fluid are $\veqv=\order{\vcv}$, and that the fluid areas of $\Scvcir$ and $\Scvend$ have areas of $\order{\rcv\lcv}$ and $\order{\rcv^2}$, respectively, the magnitude of this term is given by
\begin{equation}
\sum_{k=1}^{N+1} \eqT_{2,k} = \order{(\tau-\taucap)\rho \vcv^3 \frac{(\rcv^2 + \lcv\rcv)}{\Xcv\lcv}} = \order{\rho \vcv^2 \rcv} 
\label{eq:eqT22}
\end{equation}
In deriving the final term in this expression we have used $\order{\lcv}=\order{\rcv}$ and neglected a term containing $\taucap/\tau$ relative to one that does not, consistent with equation (\ref{eq:taucapdtau}).

\subsubsection{Examining term $\sum_{k=1}^{N+1} \eqT_{3,k}$}

This term represents the transport of gravitational potential energy into the CV during the equilibrium stages, and is given by
\begin{equation}
\sum_{k=1}^{N+1} \eqT_{3,k} = \frac{1}{\Acv} \sum_{k=1}^{N+1} \int_{\eqDtk} \int_{\Scv} \ncv \cdot \rho \hat{\Phi} ( \vvcv - \vv )  dS dt \label{eq:eqT31}
\end{equation}
In a very similar fashion to the last term this term has a magnitude given by
\begin{equation}
\sum_{k=1}^{N+1} \eqT_{3,k} = \order{(\tau-\taucap)\rho \vcv g \frac{\rcv^2}{\Xcv}} = \order{\rho g \rcv^2} 
\label{eq:eqT32}
\end{equation}
where we have additionally assumed that $\hat{\Phi} = \order{g\rcv}$ over $\Scvcir$ and $\hat{\Phi} = \order{g\lcv}$ over $\Scvend$.

\subsubsection{Examining term $\sum_{k=1}^{N+1} \eqT_{4,k}$}

The next term is a contributor to the work that the interfaces outside the CV do on the material inside the CV.  Using equations (\ref{eq:T4}) and (\ref{eq:eqT}) this term is given by
\begin{equation}
\sum_{k=1}^{N+1} \eqT_{4,k} = - \frac{1}{\Acv} \sum_{k=1}^{N+1} \int_{\eqDtk} \int_{\Scv} \sum_{i<j} \sigmaij \deltaSij \nSij \nSij : \vv \ncv  dS dt \label{eq:eqT41}
\end{equation}
As the integral contains the surface delta function $\deltaSij$, like $\eqT_{1,k}$, only four component surfaces of $\Scv$ give non-zero contributions to the integral:  Namely $\Scvtop$, $\Scvbl$, $\Scvbr$ and $\Scvend$.  We evaluate $\nSij \nSij : \veqv \ncv$ on each of these surfaces.

On $\Scvtop$, $\ncv \cdot \nSij = 0$ as here the fluid interface is perfectly flat and normal to the CV boundary (as discussed in section \ref{sec:physical_system}) so this surface makes no contribution to equation (\ref{eq:eqT41}).  For $\Scvbl$ and $\Scvbr$ the interfaces associated with these surface areas have $\vv=0$ (the solid velocity) at the CV circumference, so these surfaces also make no contribution to $\eqT_{4,k}$.  Finally, over $\Scvend$, at distances from the TPCL that are much larger than $\order{\hrough}$, $\nSij \cdot \ncv=0$ as like over $\Scvtop$ here the fluid interface is flat and normal to the CV boundary.  However at distances from the TPCL on $\Scvend$ that are of $\order{\hrough}$, $\nSij \cdot \ncv$ is not zero as in this region the interface has curvature of $\order{1/\hrough}$, as described in section \ref{sec:physical_system}.  Further, within this region the fluid velocity $\veqv = \order{\vcv}$ as the TPCL may be advancing over the solid here during these equilibrium stages.  Hence there is a contribution to the integral in equation (\ref{eq:eqT41}) from the inner part of $\Scvend$ only, giving overall
\begin{equation}
\sum_{k=1}^{N+1} \eqT_{4,k} = \order{(\tau-\taucap)\sigma\vcv\frac{\hrough}{\Xcv \lcv}} = \order{\sigma\frac{\hrough}{\lcv}}
\label{eq:eqT42}
\end{equation}
where as per equation (\ref{eq:taucapdtau}) a term involving the factor $\taucap/\tau$ has been neglected.

\subsubsection{Examining term $\sum_{k=1}^{N+1} \eqT_{5,k}$}

This term represents the work that the material stresses $\tens[M]{T}$ are doing on the material inside the CV.  Using equations (\ref{eq:T5}) and (\ref{eq:eqT}) this term is defined as
\begin{equation}
\sum_{k=1}^{N+1} \eqT_{5,k} = \frac{1}{\Acv} \sum_{k=1}^{N+1} \int_{\eqDtk} \int_{\Scv} \tens[M]{T}:\vv \ncv dS dt \label{eq:eqT51}
\end{equation}
Noting that within the solid $\vv=\vect{0}$ (as discussed section \ref{sec:material_dynamics}), and applying the fluid Newtonian stress equation (\ref{eq:newtonian_stress}) this term can be written as
\begin{align}
\sum_{k=1}^{N+1} \eqT_{5,k} & = - \frac{1}{\Acv} \sum_{k=1}^{N+1} \int_{\eqDtk} \int_{\Scvfluid} \eqp \veqv \cdot \ncv dS dt \nonumber \\
& + \frac{1}{\Acv} \sum_{k=1}^{N+1} \int_{\eqDtk} \int_{\Scvfluid} \mu \left [ \eqnablav + (\eqnablav)^T \right ] : \veqv \ncv dS dt \label{eq:eqT52}
\end{align}
where we have used the identity $\tens{I} : \veqv\ncv = \veqv \cdot \ncv$ and $\Scvfluid$ represents the surface of the CV within the fluid phase.  Equilibrium stage fluid properties $\eqp$ and $\eqnablav$ are relevant during these time intervals and are substituted from the steady-state order of magnitude expressions of equations (\ref{eq:pressure2}) and (\ref{eq:gradv}), respectively, as discussed in section \ref{sec:material_dynamics}, giving
\begin{multline}
\sum_{k=1}^{N+1} \eqT_{5,k} = - \frac{(\tau-\taucap)}{\Xcv\lcv} \times \\ \int_{\Scvfluid} \left [ \order{ \rho \vcv^2 + \frac{\mu \vcv}{\max(r,\hmol)} + \frac{\sigma}{\hsurround} } + p_0 \right ] \veqv \cdot \ncv dS \\
+ \frac{(\tau-\taucap)}{\Xcv\lcv} \int_{\Scvfluid} \order{\frac{\mu \vcv}{\max(r,\hmol)}} : \veqv \ncv dS \label{eq:eqT53}
\end{multline}
For the term involving the reference pressure $p_0$, as this is constant at any given time it can come out of the integral, leaving $\veqv \cdot \ncv$ which is zero when integrated over $\Scvfluid$ as the fluid is incompressible (using Gauss' theorem).  Hence this reference pressure term does not contribute to $\eqT_{5,k}$.  For the remainder we gather like terms and evaluate in an order of magnitude sense over the circumference and ends of the CV that are within the fluid region as
\begin{align}
%
\sum_{k=1}^{N+1} \eqT_{5,k} & = \mathcal{O} \left \{ (1-\frac{\taucap}{\tau}) \frac{1}{\lcv} \left [ \int_{\Scvcir} \left ( \rho \vcv^2 + \frac{\mu \vcv}{\max(r,\hmol)} + \frac{\sigma}{\hsurround} \right ) dS \right . \right . \nonumber \\ 
& \quad \quad \left . \left . + \int_{\Scvend} \left ( \rho \vcv^2 + \frac{\mu \vcv}{\max(r,\hmol)} + \frac{\sigma}{\hsurround} \right ) dS \right ] \right \} \nonumber \\
& = \mathcal{O} \left \{ (1-\frac{\taucap}{\tau}) \frac{1}{\lcv} \left [ \left ( \rho \vcv^2 + \frac{\mu \vcv}{\rcv} + \frac{\sigma}{\hsurround} \right ) \rcv \lcv \right . \right . \nonumber \\ 
& \quad \quad \left . \left . + \rho \vcv^2 \rcv^2 + \int_0^{\hmol} \frac{\mu \vcv}{\hmol} r dr + \int_{\hmol}^{\rcv} \frac{\mu \vcv}{r} r dr + \frac{\sigma \rcv^2}{\hsurround} \right ] \right \} \nonumber \\
& = \order{ \rho \vcv^2 \rcv } + \order{ \mu \vcv } + \order{ \frac{\sigma \rcv}{\hsurround} }
\label{eq:eqT54}
\end{align}
where for the last line we have used $\order{\rcv}=\order{\lcv}$ and employed equations (\ref{eq:lengthscales}) and (\ref{eq:taucapdtau}) to neglect comparatively small terms.

\subsubsection{Examining term $\sum_{k=1}^{N+1} \eqT_{6,k}$ \label{sec:eqT6}}

This final term required for the evaluation of equation (\ref{eq:balbar3}) represents the rate of dissipation occurring within the CV during the equilibrium periods, and is evaluated in a very similar manner to $\eqT_{5,k}$.  Using equations (\ref{eq:T6}) and (\ref{eq:eqT}) this term is defined as
\begin{equation}
\sum_{k=1}^{N+1} \eqT_{6,k} = - \frac{1}{\Acv} \sum_{k=1}^{N+1} \int_{\eqDtk} \int_{\Vcv} \tens[M]{T}:\vnabla \vv \ncv dV dt \label{eq:eqT61}
\end{equation}
Noting again that within the solid $\vv=\vect{0}$, and applying the fluid Newtonian stress equation (\ref{eq:newtonian_stress}) we find
\begin{align}
\sum_{k=1}^{N+1} \eqT_{6,k} & = \frac{1}{\Acv} \sum_{k=1}^{N+1} \int_{\eqDtk} \int_{\Vcvfluid} \eqp \tens{I} : \eqnablav dV dt \nonumber \\
& - \frac{1}{\Acv} \sum_{k=1}^{N+1} \int_{\eqDtk} \int_{\Vcvfluid} \mu \left [ \eqnablav + (\eqnablav)^T \right ] : \eqnablav dV dt \label{eq:eqT62}
\end{align}
For the pressure term we use $\tens{I} : \eqnablav = \vnabla \cdot \veqv = 0$ as the fluid is incompressible, removing this entire integral.  Substituting the equilibrium stage fluid velocity gradient from equation (\ref{eq:gradv}) into the remaining dissipation integral and evaluating in an order of magnitude sense leads to
\begin{align}
\sum_{k=1}^{N+1} \eqT_{6,k} & = \order{ \frac{\mu(\tau-\taucap)}{\Acv} \int_{\Vcvfluid} \frac{\vcv^2}{\left [ \max(r,\hmol) \right ]^2} dV } \nonumber \\
& = \orders{ \mu\vcv \left ( 1 - \frac{\taucap}{\tau} \right ) \left ( \int_0^{\hmol} \frac{r}{\hmol^2} dr + \int_{\hmol}^{\rcv} \frac{1}{r} dr \right ) } \nonumber \\
& = \orders{ \mu \vcv \ln \left (\frac{\rcv}{\hmol} \right ) }
\label{eq:eqT63}
\end{align}
where again relatively small terms have been neglected via equations (\ref{eq:lengthscales}) and (\ref{eq:taucapdtau}).

\subsubsection{Final contact angle mechanical energy balance \label{sec:finalcameb}}

With all relevant terms defined, we substitute equations (\ref{eq:totT02}), (\ref{eq:dsT04}), (\ref{eq:eqT16}), (\ref{eq:eqT22}), (\ref{eq:eqT32}), (\ref{eq:eqT42}), (\ref{eq:eqT54}), (\ref{eq:eqT63}) into equation (\ref{eq:balbar3}) giving 
\begin{multline}
\order{\rho\vcv^2 \hrough} + \order{\rho g \lcv \hrough} + \order{\frac{\sigma \hrough}{\Xcv}} = \\
\order{\rho \vcv^2 \hrough} + \order{\rho g \hrough^2} + \dsDeltasigma
- \scali[12]{\sigma} \cos \thetaa + \rightsigma - \leftsigma \\
+ \order{\rho \vcv^2 \rcv}
+ \order{\rho g \rcv^2} 
+ \order{\sigma\frac{\hrough}{\lcv}} \\
+ \order{\rho \vcv^2 \rcv} + \order{ \mu \vcv } + \order{ \frac{\sigma \rcv}{\hsurround} }
+ \orders{ \mu \vcv \ln \left (\frac{\rcv}{\hmol} \right ) }
\label{eq:barbal4}
\end{multline}
Gathering like terms and neglecting any terms that are relatively small due to the separation of lengthscales equation (\ref{eq:lengthscales}) yields
\begin{multline}
\dsDeltasigma - \scali[12]{\sigma} \cos \thetaa + \rightsigma - \leftsigma = \\
+ \order{\rho \vcv^2 \rcv}
+ \order{\rho g \rcv^2} 
+ \orders{ \mu \vcv \ln \left (\frac{\rcv}{\hmol} \right ) }
\label{eq:balbar5}
\end{multline}
Equivalently this contact angle energy balance can be expressed as
\begin{equation}
\scali[12]{\sigma} \cos \thetaa = \dsDeltasigma + \rightsigma - \leftsigma
\label{eq:balbar6}
\end{equation}
under conditions where the following inequalities hold
\begin{equation}
\order{\frac{\rho \vcv^2 \rcv}{\sigma}},
\order{\frac{\rho g \rcv^2}{\sigma}}, 
\orders{ \frac{\mu \vcv}{\sigma} \ln \left (\frac{\rcv}{\hmol} \right ) } \ll 1
\label{eq:balconditions1}
\end{equation}


\subsubsection{Interpretation of specific dissipation event surface energy change $\dsDeltasigma$ \label{sec:specificenergychange}}

The contact angle energy balance given in equation (\ref{eq:balbar6}) involves the specific dissipation event surface energy change $\dsDeltasigma$, which as previously defined is the sum of all surface energy changes occurring within the CV as a result of individual dissipation events (equation (\ref{eq:dsDeltasigma})).  Consistent with the physical model of how the contact line advances over the rough solid as outlined in section \ref{sec:material_dynamics}, surface energy is dissipated to heat (via viscous stresses) during each dissipation event resulting in a negative $\dsDeltasigma$.  Hence an alternative nomenclature for this term that is consistent with this contact line movement model (as discussed in the introduction, \citep[e.g.][]{joanny84}) is to define the specific dissipation per projected area of solid traversed due to surface roughness as $D \approx -\dsDeltasigma$.  In this section we formalise this relationship by performing an order of magnitude analysis on an energy balance conducted over all dissipation events occurring during the advance period to show that the surface energy liberated via $\dsDeltasigma$ is in fact dissipated to heat during these periods.

In order to perform this dissipation event analysis, we need models for how the materials within the CV behave during the dissipation periods.  As for the equilibrium periods, we assume that the solid does not deform (following the justifications from section \ref{sec:material_dynamics}) giving $\vv=0$ within the solid during these times.  For the fluids we require a description of the velocity $\vdsv$, velocity gradient $\dsnablav$ and pressure $\dsp$ existing during these times.  For the velocity, noting that the continuous equilibrium velocities still exist within the dissipation periods, and assuming that significant capillary driven velocities only exist within an region of size $\order{\hrough}$ local to each dissipation event (as the Reynolds number of these motions is not large), we define
\begin{equation}
\vdsv (\vect{x}) = 
\begin{cases}
\veqv + \order{\vcap} & \text{if } \rcap<\order{\hrough} \\
\veqv & \text{otherwise}
\end{cases}
\label{eq:dsv}
\end{equation}
where $\rcap=|\vect{x}-\xcap|$ is the distance to the centre of the relevant $k$th dissipation event, and $\xcap$ is the location of the particular $k$th dissipation event that is centred on the centreline of the CV \footnote{For notational simplicity here and in subsequent dissipation event model definitions we do not indicate what specific dissipation event variables such as $\vdsv$, $\xcap$ and $\rcap$ refer to}.  For the velocity gradient we similarly define
\begin{equation}
\dsnablav (\vect{x})= 
\begin{cases}
\eqnablav + \order{\frac{\vcap}{\max(\rcap,\hmol)}} & \text{if } \rcap<\order{\hrough} \\
\eqnablav & \text{otherwise}
\end{cases}
\label{eq:dsnablav}
\end{equation}
where as per the equilibrium velocity gradient model of equation (\ref{eq:gradv}) we limit the stress generated at the moving TPCL that is within $\order{\hmol}$ of the solid.  Finally for pressure we perform another order of magnitude analysis on the Navier-Stokes equations, but now recognise that capillary induced pressure changes occur over the interface that is deforming over length $\order{\hrough}$ during these times, giving
\begin{equation}
\dsp (\vect{x})= 
\begin{cases}
\eqp + \order{\rho \vcap^2} + \order{\frac{\mu\vcap}{\max(\rcap,\hmol)}} + \order{\frac{\sigma}{\hrough}} & \text{if } \rcap<\order{\hrough} \\
\eqp & \text{otherwise}
\end{cases}
\label{eq:dsp}
\end{equation}
This completes the dissipation event material specifications.

The dissipation event energy analysis now largely mirrors that conducted for the equilibrium stages (as detailed in sections \ref{sec:eqT1} to \ref{sec:eqT6}) by summing the individual dissipation event energy balances of equation (\ref{eq:balhat2}) over all $N$ dissipation events.  The details of this analysis are contained within Appendix \ref{sec:dissipationordermag}, with the final conclusion of the energy balance being
\begin{align}
\dsDeltasigma & = \order{\frac{\taucap}{\tau} \sum_{i=1}^6 \sum_{k=1}^{N+1} \eqT_{i,k} } + \order{\rho \vcv^2 \hrough} + \order{\rho g \hrough^2} \nonumber \\
& + \order{\frac{\rho \vcap^2 \hrough^2}{\lcv}} + \order{\frac{\rho g \hrough^3}{\lcv}} + \order{\sigma \frac{\hrough}{\lcv}} \nonumber \\
& + \order{ \mu \vcap \frac{\hrough}{\lcv}} + \orders{ \mu \vcap \ln \left ( \frac{\hrough}{\hmol} \right ) }
\label{eq:dsbal2}
\end{align}
Under conditions for which the inequalities of equation (\ref{eq:balconditions1}) hold, the equilibrium terms represented by $\sum_{k=1}^{N+1} \eqT_{i,k}$ in the above have a maximum magnitude of $\order{\sigma}$, which is the same as that of $\dsDeltasigma$.  As $\order{\frac{\taucap}{\tau}} \ll 1$ via equation (\ref{eq:taucapdtau}), it follows that the equilibrium terms make no significant contribution to equation (\ref{eq:dsbal2}) and can be neglected.  For the remaining terms, using the lengthscales assumption equation (\ref{eq:lengthscales}) in combination with equation (\ref{eq:balconditions1}) yields after simplifications
\begin{equation}
\dsDeltasigma = \order{\frac{\rho \vcap^2 \hrough^2}{\lcv}} + \orders{ \mu \vcap \ln \left ( \frac{\hrough}{\hmol} \right ) }
\label{eq:dsbal3}
\end{equation}
 
Tracing back through the dissipation energy analysis presented in Appendix \ref{sec:dissipationordermag} shows that the first and second terms on the RHS of equation (\ref{eq:dsbal3}) represent the transport of kinetic energy (and associated pressure work) through the ends of the CV due to dissipation event velocities, and the viscous dissipation of energy that is converted to heat around the moving TPCL during dissipation events, respectively.  The ratio of these two terms depends on the roughness scale capillary velocity Reynolds number as well as various lengthscale ratios, so that in general neither can assumed to be dominant.  The second term directly represents energy conversion to heat during the capillary driven dissipation events, as envisaged.  While the first term does not directly relate to energy dissipation, it does represent the transport of energy through the ends of the CV and hence parallel to the TPCL.  As the analysis shows that no significant energy transport occurs over the circumference of the CV during the dissipation events, the energy represented by the first term in equation (\ref{eq:dsbal3}) remains within the vicinity of the contact line, and so will eventually be dissipated to heat via the second term in the same equation, but in an adjacent control volume.  Hence, overall equation (\ref{eq:dsbal3}) shows that the energy liberated by surface changes that occur during each dissipation event is dissipated to heat within the vicinity of the TPCL, and we are justified in using the nomenclature $D=-\dsDeltasigma$ to represent the specific energy dissipation per area travelled due to surface roughness.

It should be reiterated however that the discussion in this section does not change the way in which $D$ (or equivalently $-\dsDeltasigma$) is evaluated, being defined solely in terms of surface energy changes that occur over all dissipation events.  Indeed a strength of this framework is that $D$ can be calculated based solely on these surface energy changes, rather than needing to know the dynamic and transient details of each individual dissipation event.

\section{Discussion \label{sec:discussion}}

\subsection{Summary of key results \label{sec:summary}}

The contact angle mechanical energy balance equations are summarised in table \ref{tab:summary}.  The requirements of inequality equation (\ref{eq:balconditions1}) have been combined with previous lengthscale assumptions and interpreted as conditions on $\rcv$ and $\vcv$.  $\order{\lcv}=\order{\rcv}$ is applied throughout. The limiting CV size variable $\rcvgrav$ is the capillary length that in this analysis has originated from limiting the transport of gravitational potential energy during the equilibrium stages of the energy balance.  The three limiting control volume velocities $\vcvke$, $\vcvvis$ and $\vcvcap$ have originated from limiting the transport of kinetic energy through the CV boundaries during the equilibrium stages, limiting the rate of viscous dissipation occurring at a molecular level around the moving TPCL during the equilibrium stages, and ensuring that the duration of dissipation events occurring within the CV is small compared to the duration of interface movement, respectively.  As per section (\ref{sec:specificenergychange}), the specific change in surface potential energy occurring during the dissipation periods has been defined in terms of $D$, being the specific dissipation due to roughness.

\begin{table}
\center
\begin{framed}
\begin{minipage}{0.95\textwidth}
\begin{equation}
\scali[12]{\sigma} \cos \thetaa = \rightsigma - \leftsigma - D
\label{eq:mebfinal}
\end{equation}
subject to
\begin{gather}
\hmol \ll \hrough \ll \rcv \ll \hsurround , \rcvgrav \label{eq:lengthscales2} \\
\vcv \ll \min \left ( \vcvke ,\vcvvis ,\vcvcap \right ) \label{eq:vcvconditions}
\end{gather}
where
\begin{gather}
\leftsigma = \sum_{i<j} \sigmaij \frac{\leftareaij}{\Acv} \quad \text{and} \quad \rightsigma = \sum_{i<j} \sigmaij \frac{\rightareaij}{\Acv} \tag{\ref{eq:areasigmas}}\\
D = -\sum_{k=1}^N \sum_{i<j} \sigmaij \frac{\widehat{\Delta \Aij}_k}{\Acv} \label{eq:D}\\
\vcap = \min \left ( \frac{\sigma}{\mu}, \sqrt{\frac{\sigma}{\rho\hrough}} \right ) \tag{\ref{eq:vcap}}\\
\vcvke = \sqrt{ \frac{\sigma}{\rho \rcv}} \label{eq:vcvke}\\
\vcvvis = \sqrt{ \frac{\sigma}{\mu \ln \left (\rcv/\hmol \right )}} \label{eq:vcvvis}\\
\vcvcap = \vcap \frac{\hrough}{\rcv} \label{eq:vcvcap}\\
\rcvgrav = \sqrt{\frac{\sigma}{\rho g}} \label{eq:rcvgrav}
\end{gather}
\end{minipage}
\end{framed}
\caption{Equation summary for the contact angle mechanical energy balance.  $D$, $\protect\leftsigma$, $\protect\rightsigma$ are the specific dissipation due to roughness, leaving compound solid surface energy and entering compound solid surface energy, respectively.}
\label{tab:summary}
\end{table}

It is illustrative to consider some simple examples of the application of equation (\ref{eq:mebfinal}).  In each of these cases phase \circleme{1} is advancing to the right at $\thetaa$ while displacing phase \circleme{2} (as per Figure \ref{fig:three_phase}):
\begin{itemize}
\item Youngs: For a flat surface the leaving compound surface energy is $\leftsigma=\scal[1s]{\sigma}$, the entering compound surface energy is $\rightsigma=\scal[2s]{\sigma}$ and the specific roughness dissipation rate $D=0$ as no dissipation events occur during the TPCL advance.  Application of equation (\ref{eq:mebfinal}) gives $\scali[12]{\sigma} \cos \thetaa = \scal[2s]{\sigma} - \scal[1s]{\sigma}$, or via Young's equation, $\thetaa=\thetae$.
\item Wenzel: Here the leaving surface energy is $\leftsigma=r\scal[1s]{\sigma}$ and the entering surface energy is $\rightsigma=r\scal[2s]{\sigma}$, where $r$ is the roughness or specific solid area (total solid area per projected solid area).  Application of equation (\ref{eq:mebfinal}) gives $\cos \thetaa = r \cos \thetae - D/\scali[12]{\sigma}$ which is consistent with Wenzel's equation, but augmented by a roughness dissipation term.  Here increasing the roughness $r$ can either increase or decrease the advancing angle depending on $\thetae$, however increasing the roughness dissipation $D$ always increases the contact angle.
\item Cassie: In this general case $f_1$ and $f_2$ are defined as the wetted and non-wetted liquid areas under the droplet per projected area, respectively, with $r$ defined as per the Wenzel case.  Under these conditions the leaving surface energy is $\leftsigma=-f_1 \scal[12]{\sigma}\cos \thetae+f_2 \scal[12]{\sigma}+r\scal[2s]{\sigma}$ and the entering surface energy is $\leftsigma=r\scal[2s]{\sigma}$.  Substituting into equation (\ref{eq:mebfinal}) gives $\cos \thetaa = f_1 \cos \thetae - f_2 - D/\scali[12]{\sigma}$ which is consistent with Cassie's analysis\citep{cassie44}, but augmented with a dissipation term.  For a droplet resting on the top of flat structures (for example `Fakir' droplets existing on photolithography based surfaces) this equation can be further simplified using $f_2 = 1 - f_1$.
\end{itemize}

In general the specific roughness dissipation $D$ and leaving compound surface energy $\rightsigma$ cannot be guessed from the solid topology alone and instead must be modelled from an understanding of the fluid dynamics occurring within the CV, or measured.  The entering compound surface energy $\leftsigma$ depends on the history of the surface and must similarly be modelled or measured.  Hence the mechanical energy balance equation (\ref{eq:mebfinal}) is best viewed as a framework within which results from dynamic interfacial modelling or measurements can be incorporated for specific combinations of solid topologies and fluid phases.  These activities are the topic of on-going work in our lab.

In terms of calculating the receding angle, as per Figure \ref{fig:three_phase} when phase \circleme{1} is advancing over the solid, phase \circleme{2} is receding.  Hence substituting $\thetaa=\pi-\thetar$ into equation (\ref{eq:mebfinal}) and letting the compound surface energies under phases \circleme{1} and \circleme{2} while phase \circleme{2} is receding equal $\scal[r]{\rightsigma}=\leftsigma$ and $\scal[r]{\leftsigma}=\rightsigma$, respectively, yields $\scali[12]{\sigma} \cos \thetar = \scal[r]{\rightsigma} - \scal[r]{\leftsigma} + \scal[r]{D}$, with $\scal[r]{D}$ now referring to the dissipation occurring while phase \circleme{2} recedes.  As expected an increase in $\scal[r]{D}$ leads to a decrease in $\thetar$.

\subsection{Range of validity \label{sec:validity}}

\begin{figure}[h]
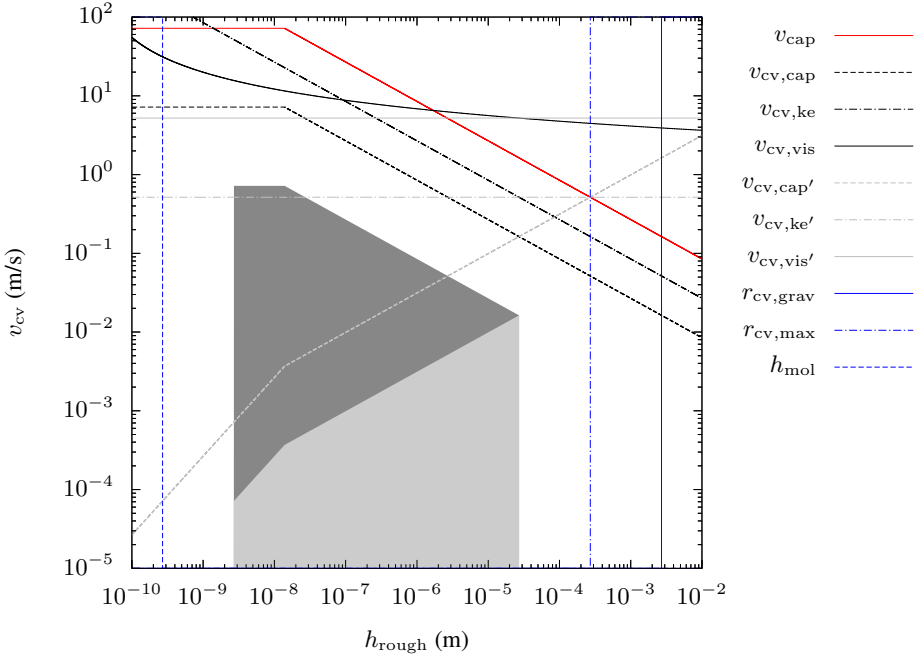%
\caption{Regions of validity (shaded) for the contact angle energy analysis based on a typical water/air system.  The velocity limits $\vcvcap$, $\vcvke$ and $\vcvvis$ and the entire shaded validity region (dark grey plus light grey) assume that the CV size depends on the solid roughness according to $\rcv/\hrough=10$.  Conversely, the velocity limits $\vcvcapp$, $\vcvkep$ and $\vcvvisp$ and light grey validity region assume that the CV size is constant at the maximum allowable value of $\rcv=\rcvmax=\rcvgrav/10$.}
\label{fig:conditions}
\end{figure}

One strength of this energy balance approach for predicting contact angles is that it not only provides a framework for predicting the angle, but also specifies under what conditions the predicted angle will be valid.  Specifically, for equation (\ref{eq:mebfinal}) to be representative of a real system all conditions specified by equations (\ref{eq:lengthscales2}) and (\ref{eq:vcvconditions}) must be satisfied.  Continuing the discussion from section \ref{sec:material_dynamics}, we examine what physical limitations these conditions place on applying the theory to the common water/air system, with results shown in Figure \ref{fig:conditions}.

In terms of lengthscales, evaluating $\rcvgrav$ from equation (\ref{eq:rcvgrav}) requires that $\rcv \ll \rcvgrav = 2.7\units{mm}$, which is approximately satisfied if we adopt $\rcv=\rcvmax=\rcvgrav/10=0.27\units{mm}$ for the water system (that is, one order of magnitude less).  From the separation of lengthscales equation (\ref{eq:lengthscales2}) this places an upper constraint on the roughness applicable under the theory of $\hrough \lessapprox 27\units{\mu m}$, which is indicated by the right bound of the shaded area in Figure \ref{fig:conditions}.  Hence, gravitational effects place an upper limit on the size of roughness applicable under this theory.  At the small scales we also require that $\hmol \ll \hrough$, and given that the molecular size of water is around $0.27\units{nm}$ this places a lower limit on $\hrough$ of around $2.7\units{nm}$ as indicated by the left bound of the shaded area in Figure \ref{fig:conditions}.  Note that some theories use a larger length cut-off of $100\units{nm}$ in their analysis citing the influence of van der Waals and double-layer forces at these lengthscales\citep{de-gennes85}:  While still continuum forces, these are not accounted for in our hydrodynamically-based framework.  Non-continuum effects such as thermal fluctuations that could influence interface topologies at very small lengthscales are also not considered.  Hence, while the lower roughness size limit of this theory requires further validation, overall the conclusion is that CAH can be significant even for surfaces that have very small sized roughness (at least in an engineering sense), a conclusion that is supported by experiments\citep{delmas11,fetzer11}.

It is interesting that aside from these lengthscale and velocity constraints (discussed below), there is no reference to the absolute size of the surface roughness contained within the energy analysis.  While models of $D$, $\leftsigma$ and $\rightsigma$ may in some cases depend on the absolute size of the surface roughness, it is likely that for many fluid combinations and solid topologies the advancing and receding contact angles do not depend on the absolute size of the roughness, but only on the topology of the roughness.  As discussed in the introduction, this is consistent with a growing number of observations \citep{oner00,dorrer08,li16a,jiang19}.

In terms of the speed that the CV can travel at for the theory still to be valid (that is, $\vcv$), the constraints partly depend on how the size of the CV is defined.  We consider two scenarios:  In the first the CV size decreases with $\hrough$ such that $\rcv/\hrough=10$;  In the second the CV size is kept constant at the maximum value of $\rcv=\rcvmax=0.27\units{mm}$ as determined by equation (\ref{eq:lengthscales2}).  The various limiting velocities corresponding to the first decreasing CV size case are indicated by $\vcvcap$, $\vcvke$ and $\vcvvis$ in Figure \ref{fig:conditions}, while the limiting velocities corresponding to the second constant CV size case are indicated by $\vcvcapp$, $\vcvkep$ and $\vcvvisp$.  Similarly, the ranges of theory validity are shown in the figure by the entire shaded region for the first case, and by the light shaded region for the second case.  In practice the CV size determines at what distance from the solid (or TPCL) we are measuring the `macroscopic' advancing angle, $\thetaa$.

For both CV size cases considered the capillary velocity constraint, which is an interpretation of equation (\ref{eq:taucapdtau}) and requires that the total dissipation event time during the advance ($\taucap$) is less than the total advance time ($\tau$), is the only constraint that limits the advance velocity over the entire range of applicable roughness scales.  As was discussed in section \ref{sec:material_dynamics} and shown in Figure \ref{fig:conditions}, the capillary velocity $\vcap$ for this water system increases as $\hrough$ decreases, reaching $\sim 72 \units{m/s}$ below $\hrough = \hroughcrit \approx 14\units{nm}$.  For the first decreasing CV size case this velocity constraint allows $\vcv$ to remain at practically large values (of $\sim 0.72\units{m/s}$) even at the lowest limit of applicable $\hrough$.  For the second constant CV size case however the capillary velocity constraint decreases the viable $\vcv$ as $\hrough$ decreases, reaching $\sim 70 \units{\mu m/s}$ at the lowest applicable $\hrough \approx 2.7 \units{nm}$.

It is notable that for most roughness sizes and under either CV size scenarios the velocity of the interface under which the theory is valid, and hence predicts that the advancing angle will be equal to the static advancing angle of the system, is in practical terms large.  At the highest valid solid roughnesses that are in the tens of micron range the advancing angle will remain as the static angle until the interface is moving faster than several cm/s.  At the lowest valid roughnesses the maximum valid advance velocity depends on the choice of CV size, but even for the more restrictive larger CV size case the maximum advance velocity is still around $0.1\units{mm/s}$ at the smallest roughnesses, which is certainly an experimentally accessible value for performing contact angle measurements.  In general these results emphasise that contact angle hysteresis depends more on surface roughness than interface velocity, for most practical surfaces experiencing interface movement for most engineering applications.

It is also interesting that the maximum allowable interface speed is in general a function of the CV size, or put differently, that the predicted dynamic contact angle is a function of the distance from the TPCL (and hence solid) that the angle is measured.  This has been noted previously in the context of dynamic contact angle models that are based on the concept of viscous dissipation \citep{de-gennes85}.  As previously discussed, for the second CV size regime as specified by the light grey area in Figure \ref{fig:conditions}, the upper velocity constraint is determined by $\vcvcapp$, which can alternatively be interpreted as a constraint on the maximum number of solid defects existing per solid area over which the CV advances ($\Acv$).  This constraint is represented by equation (\ref{eq:taucapdtau}).  However, it is possible that this constraint could be relaxed if a different approach to the energy balance was adopted:  Rather than requiring that the total time for dissipation events be relatively small, an alternative approach that was advanced in earlier versions of this theory \citep{dhasc08} is to locate dissipation events not only in time but also in space within the CV, allowing more than one dissipation event to be occurring within the analysis concurrently.  This approach is mathematically more onerous to produce rigorous conditions on theory validity, but is worthy of future research as it could relax the $\vcvcap$ constraint on $\vcv$ under conditions where the CV size is of a macroscopic (i.e. mm) size but the solid roughness much smaller.  Note that if the $\vcvcap$ restraint were relaxed, the next constraint on the dynamic contact angle would come from $\vcvke$ which is related to kinetic energy transport through the CV volume, rather than $\vcvvis$ which is concerned with viscous dissipation occurring within the CV volume (and more specifically around the TPCL).  Interestingly previous theories have focused on predicting dynamic contact angles from knowledge of viscous or molecular energy dissipation occurring around the TPCL\citep{moffatt64,huh71,voinov76,de-gennes85,petrov92}, rather than considering kinetic energy transport within the TPCL region --- in contrast to this framework that suggests that kinetic energy transport is more important than viscous dissipation in predicting dynamic contact angles.  Validating this result and producing new models of dynamic contact angle is another area for future research suggested by this energy conservation framework.

\subsection{Other model limitations \label{sec:limitations}}

In addition to the advance velocity, roughness size and CV size constraints discussed in the previous section, other assumptions used in the derivation of the energy balance theory will not be valid for some systems, but could form avenues for future extensions of the theory.

Inherent in the contact angle framework, and enabling multiphase mechanical energy balance presented in section \ref{sec:mmeb}, is the assumption that there is no irreversible work involved in interface creation (or destruction).  Physically irreversibilities could be present in a system via chemical or molecular effects associated with interface formation, including effects due to surfactants or other types of surface active molecules.  In the context of dynamic contact angle modelling, some work has been done in incorporating molecular dissipation and adsorption when predicting advancing contact angles\citep{blake69,de-gennes85,brochard-wyart92,mohammad-karim22}.  Following this work and recognising the parallels between these interfacial formation irreversibilities and dissipation due to TPCL movement, it is likely that this irreversible interfacial work will appear in the contact angle energy balance as a dissipation term that is in addition to the hydodynamically based $D$, but this does require reformulation of the general mechanical energy balance and application to the moving contact line CV to show this rigorously.  A further assumption used by the model is that the solid is chemically homogeneous.  This assumption could be relaxed in a straight-forward manner by including more than one solid interface type in the contact angle energy analysis.

Another limitation of the presented work is that fluids are assumed to be incompressible, and no dissolution or evaporation of the fluids is permitted.  Both of these assumptions are physically limiting when one of the phases is a gas.  Physically a gas will behave differently to a liquid in cases where a micro-bubble is formed within the solid roughness, caused by the specific dynamics of the fluid interface as it advances over solid defects.  Within a formed micro-bubble the laplace pressure will be high ($\order{\sigma/\hrough}$) which will cause the density of gas to increase and hence the volume of the micro-bubble to decrease.  How significant these affects are depends on the bubble size, interfacial tension and gas equation of state.  Further, high pressures within the bubble will drive gas dissolution into the surrounding fluid phase, further reducing the volume of the formed micro-bubbles.  These changes to the micro-bubble size will in turn affect the leaving compound solid surface energy $\leftsigma$, affecting the predicted advancing contact angle.  In terms of the energy framework, fluid compressibility could be implemented quite easily by relaxing the $\vnabla \cdot \vv = 0$ constraint used when evaluating the boundary pressure work terms.  Gas dissolution is more complex however, as this process is transient, and so the framework would need to recognise over what timescale the dissolution process is taking place, and hence where in the energy balances the changes in (particularly) interfacial energy should occur.  Micro-droplet evaporation is a related phenomena not accounted for by the theory that could also change $\leftsigma$ and $\rightsigma$ depending on the wetting history of the solid surface and volatility of the fluids.  It would be interesting to account for dissolution and evaporation effects within this framework as both introduce a transient or history effect to the wetting behaviour of the system.  Again, this suggests avenues for future theoretical and experimental research.

A further limitation of the theory relates to the size of roughness considered.  As presented the roughness is characterised by a single lengthscale $\hrough$ which represents both the height of surface defects and the spacing between them.  An extension to this theory could distinguish between these two lengthscales, allowing better analysis of surfaces composed of a dilute number of strong defects, or surfaces that are gently undulating.  A second assumption relating to the size of the defects is that only one lengthscale is considered in the analysis, whereas real surfaces (particularly biological inspired super-hydrophobic surfaces) can possess a hierarchical range of roughnesses\citep{feng02}.  A potential extension of the theory would be to use a cascade of CV sizes to predict the contact angles at each lengthscale, with the equilibrium contact angles used at each lengthscale determined from contact angle predictions performed using smaller lengthscale CVs.  In this way the macroscopic contact angle could be determined from knowledge of roughness topologies at each lengthscale.

Finally the stress model for the solid material used in this study effectively implies that there are no energy changes occurring within the CV that are related to the solid phase.  More complex solid stress models could certainly be incorporated into the framework, representing (for example) the advance of fluids over soft or semi-liquid materials.  Introducing deformable solids would not only introduce additional dissipation, but would also require additional dynamical modelling or measuring of the fluid, solid and interfacial behaviours occurring around the TPCL.  This is also the topic of on-going work.

\section{Conclusion \label{sec:conclusion}}

Starting from a statement of momentum conservation, a mechanical energy conservation framework has been derived that allows the contact angle hysteresis (CAH) range to be predicted from knowledge of the interfacial dynamics that occur around an advancing three phase contact line (TPCL).  Unlike most previous works the analysis is not specific to a particular wetting regime (e.g., Cassie or Wenzel) or particular surface structure (e.g., holes, poles, periodic, dilute etc).  The analysis also resolves a number of questions about wetting on rough surfaces that have been the source of confusion in the literature:
\begin{itemize}
\item  A suitable control volume (CV) can be defined that is anchored to the TPCL around which mechanical energy conservation can be rigorously performed.  This shows that at least for the static CAH range, and conditional upon the CV lengthscale and speed satisfying four conditions (i.e., equations (\ref{eq:vcvke}-\ref{eq:rcvgrav})), the advancing and receding angles are independent of the surrounding flow.  This is significant as it means that the energy of the entire flow system (e.g., the entire drop) need not be considered when predicting CAH, resolving extensive discussion within the field\citep{gao07a,mchale07,gao07b,nosonovsky07,panchagnula07,marmur22}.
\item  By adopting semi-quantitative models for how the fluids and phase interfaces behave within the advancing CV that are based on experimental observation, we have shown from energy conservation that CAH directly results from the `pinning' and `depinning' or `jumping' of the fluid interface, deriving a general expression for this dissipation in terms of surface areas pre and post each jumping event.  No dissipation or CAH results from contact line distortion (in isolation), showing (in the language of \citet{joanny84}) that only `strong' defects (as opposed to `weak' defects) cause CAH.  This is consistent with the initial observations of \citet{joanny84} but conflicts with some subsequent studies \citep{pomeau85,robbins87,opik00}.
\item  Studies that consider how a fluid interface advances over a rough surface have previously inferred what the macroscopic CAH range is from the results, but without clear justification\citep{cox83}.  The present work shows how interfacial and energy analyses can be related:  Namely, interfacial dynamic models can be used to predict dissipation and generated (leaving) compound solid surface energies which can then be fed back into the derived energy conservation framework to predict the CAH range.  Only for specific simple systems will the CAH range inferred from interfacial analysis match that predicted via energy conservation and observed at the macroscopic scale. 
\item  While not the central focus of this study, the energy conservation analysis suggests that when calculating dynamic contact angles kinetic energy transport around the TPCL is at least as significant as hydrodynamic energy dissipation occurring around the TPCL.  This requires more theoretical and experimental confirmation, but could explain some of the discrepancies between existing dynamic contact angle models and experimental observation \citep{mohammad-karim22}.
\end{itemize}
Many limitations of the framework have also been discussed and could form the basis of future extensions, including considering compressible, evaporating or dissolving fluids, irreversible work associated with interface creation or destruction, solid chemical heterogeneity, soft solids, and non-negligible CV advance speeds.  These are topics of on-going work in our group.

\section{Nomenclature}


\begin{supertabular}{lp{10cm}}
%
\multicolumn{2}{l}{\em{Arabic:}}\\
$\scali[ij]{A}$ & area of $ij$ interface type\\
$\scali[ij]{\rightarea}$ & area of $ij$ interface type entering the control volume from the right\\
$\scali[ij]{\leftarea}$ & area of $ij$ interface type leaving the control volume on the left\\
$\Acv$ & projected area of solid surface swept by CV over advance duration\\
$\widehat{\Delta \Aij}_k$ & change in interfacial area $ij$ during dissipation period $k$\\
$D$ & total energy dissipation occurring during advance\\
$\vecti[k]{e}$ & coordinate unit vector in direction $k$\\
$\vect{g}$ & gravity vector\\
$\hrough$ & lengthscale of solid surface roughness\\
$\hroughcrit$ & lengthscale of solid surface roughness that delimits viscous and inertial fluid flow regimes\\
$\hsurround$ & lengthscale of surround flow\\
$\hmol$ & lengthscale of molecular or non-continuum effects within the fluid\\
$\tens{I}$ & identity tensor\\
$\lcv$ & length of CV\\
$\ncv$ & outward normal to surface of CV\\
$\nSij$ & unit normal vector of $ij$ interface type, directed into phase $i$\\
$m$ & number of material phases present within control volume\\
$p$ & pressure\\
$p_0$ & reference pressure\\
$\dsp$ & pressure within the fluid during a dissipation period\\
$\eqp$ & pressure within the fluid during an equilibrium period\\
$\rcv$ & radius of CV\\
$\rcap$ & distance to centre of dissipation event\\
$\rcvgrav$ & capillary length\\
$\rcvmax$ & maximum applicable CV size\\
$\Scv$ & surface of CV\\
$\Scvcir$ & circumferential surface of CV\\
$\Scvend$ & end surface of CV\\
$\Scvbr$ & circumferential surface of CV at the bottom right containing compound solid interfaces\\
$\Scvbl$ & circumferential surface of CV at the bottom left containing compound solid interfaces\\
$\Scvtop$ & circumferential surface of CV at the top containing interface type $12$\\
$\Scvfluid$ & surface of CV within fluid phases\\
$t$ & time\\
$\dstk$ & start of dissipation period $k$\\
$\eqtk$ & start of equilibrium period $k$\\
$\dsDtk$ & duration of dissipation period $k$\\
$\eqDtk$ & duration of equilibrium period $k$\\
$\totT_i$ & energy term $i$ corresponding to total analysis period\\
$\dsT_{i,k}$ & energy term $i$ corresponding to dissipation period $k$\\
$\eqT_{i,k}$ & energy term $i$ corresponding to equilibrium period $k$\\
$V$ & volume\\
$\Vcv$ & volume of CV\\
$\Vcvfluid$ & volume of CV containing fluid phases\\
$\leftV$ & volume swept by $\Scvbl$ over advance duration\\
$\rightV$ & volume swept by $\Scvbr$ over advance duration\\
$\vect{v}$ & velocity\\
$\vcap$ & speed of fluid movement during dissipation events\\
$\vcv$ & speed of control volume advance\\
$\vvcv$ & velocity of control volume advance\\
$\veqv$ & velocity of fluids during equilibrium periods\\
$\vdsv$ & velocity of fluids during dissipation periods\\
$\eqnablav$ & velocity gradient within fluids during equilibrium periods\\
$\dsnablav$ & velocity gradient within fluids during dissipation periods\\
$\Xcv$ & distance travelled by CV during total analysis period\\
$\xcv$ & distance from start of CV travel\\
$\xcap$ & location of dissipation event\\
$\vecti[\mathrm{s},ij]{x}$ & location of $ij$ interface type \\
\\ \multicolumn{2}{l}{\em{Greek:}} \\
$\delta$ & one dimensional Dirac delta function \\
$\deltaSij$ & surface delta function for $ij$ interface type\\
$\dsDeltasigma$ & sum of change in surface energies occurring over all dissipation events\\
$\scali[\mathrm{S},ij]{\delta}$ & surface delta function for $ij$ interface type \\
$\mu$ & viscosity\\
$\hat{\Phi}$ & gravitational potential function\\
$\rho$ & density\\
$\sigma$ & energy per unit area (interfacial tension)\\
$\sigmaij$ & energy per unit area (interfacial tension) of $ij$ interface (between phases $i$ and $j$)\\
$\rightsigma$ & compound energy per unit area solid surface entering the CV from the right and under phase \circleme{2}\\
$\leftsigma$ & compound energy per unit area solid surface leaving the CV on the left and under phase \circleme{1}\\
$\sum_{i<j}$ &  sum taken over all interface types ($=\sum_{j=1}\sum_{i=1}^{j-1}$)\\
$\tens[M]{T}$ & material stress tensor\\
$\tens[S]{T}$ & interfacial stress tensor\\
$\taucap$ & total duration of dissipation events occurring during CV advance\\
$\thetaa$ & macroscopic advancing angle of phase \circleme{1} in phase \circleme{2}\\
$\thetar$ & macroscopic receding angle of phase \circleme{1} in phase \circleme{2}\\
$\thetae$ & equilibrium angle of phase \circleme{1} in phase \circleme{2}\\
%
%
\end{supertabular}

\backsection[Acknowledgements]{An early version of the work was presented at the 82nd ACS Colloid and Surface Science Symposium \citep{dhasc08}.  The author acknowledges conversations and collaborations with many people over this timeframe regarding the application of this theory, including with Prof Franz Grieser, Prof Xuehua Zhang, A/Prof Brigitte Stadler, Prof William Ducker, Prof David Dunstan, Prof Paul Mulvaney, Dr Annette Haebich, Dr Srinivas Mettu, A/Prof Hong Zhao, Ms Mary Jane, Mr Pawan Kumar and Prof Suman Chakraborty.  Pawan Kumar is currently completing a PhD at the University of Melbourne on the application of this theory, supported by a Melbourne-India Postgraduate Program scholarship.}

\backsection[Funding]{The majority of this theory was developed in 2007 while the author was between employment contracts.  The work formed the basis for several unsuccessful Australian Research Council Discovery Project applications between the years of 2008 and 2017 (DP0881399, DP0988840, DP150104699, DP160103697 and DP170104610).  The author gratefully acknowledges the support of an Melbourne Institute of Materials Interdisciplinary Seed-Funding Scheme from the University of Melbourne in 2010 that was used to generate experimental results to support this theory.}

\backsection[Declaration of interests]{The authors report no conflict of interest.}

\backsection[Author ORCIDs]{D.J.E. Harvie, https://orcid.org/0000-0002-8501-1344}


\appendix

\section{Appendix: Properties of the surface delta function ($\scal[S]{\delta}$) \label{sec:deltaproperties}}

In this appendix we derive three properties of the surface delta function that are used in Section \ref{sec:mmeb} in the derivation of the multiphase mechanical energy balance.

\subsection{Volume integral of surface delta function \label{sec:deltapropertiesvolume}}

We start with some definitions:  For an interface that lies between a specific combination of immiscible materials the surface delta function is given by \citep{lafaurie94}
\begin{equation}
\scali[\mathrm{S}]{\delta}(\vect{x})=\delta(\scali[1]{q})
\label{eq:deltadef} ,
\end{equation}
where $\scali[1]{q}$ is the distance between $\vect{x}$ and the closest point on the interface surface, and $\delta$ is the one dimensional Dirac delta function.  As $\delta=0$ when $\scali[1]{q}\ne 0$, equation~(\ref{eq:deltadef}) shows that $\scal[S]{\delta}$ is nonzero only on the interface surface.

For the analysis that follows we require a more specific relationship between $\vect{x}$ and $\scali[1]{q}$ that is consistent with equation~(\ref{eq:deltadef}):  Given that $\scali[1]{q}=0$ defines the interface surface, we specify the location of any point $\vect{x}$ that is near the surface via a series of coordinates $\vect{q}=(\scali[1]{q},\scali[2]{q},\scali[3]{q})$ such that
\begin{equation}
\vect{x}(\vect{q})=\vect[S]{x}(\scali[2]{q},\scali[3]{q}) + \scali[1]{q}\vect[S]{n}(\scali[2]{q},\scali[3]{q})
\label{eq:xtoqdef} .
\end{equation}
Here $\scali[2]{q}$ and $\scali[3]{q}$ are a pair of convected surface coordinates \citep{aris62} that uniquely locate a material particle at $\vect[S]{x}$ that lies on the interface surface, and $\vect[S]{n}$ is a unit normal to the surface at $\vect[S]{x}$ which is (consistently) directed into one of the phases.  As $\vect[S]{x}$ moves with the material, when $\scali[2]{q}$ and $\scali[3]{q}$ are held constant we have
\begin{equation}
\frac{d\vect[S]{x}}{dt} =\vect{v}
\label{eq:xsvelocity} 
\end{equation}
where $\vect{v}$ is the local material velocity.

Lines of constant $\scali[2]{q}$ and $\scali[3]{q}$ define the surface coordinate lines.  As the surface coordinates move with the material, the coordinate lines will not in general be orthogonal to each other, even if they are initially.  Provided that all material strain rates remain finite however, they will not become coincident:  This is important as it means that provided $\vect[S]{x}$ can be defined uniquely in terms of $\scali[2]{q}$ and $\scali[3]{q}$ at some particular time, a unique relationship between $\vect[S]{x}$ and ($\scali[2]{q}$,$\scali[3]{q}$) will be realisable for all time (on a smooth and continuous surface).

\begin{figure}[h]
\centering
\resizebox{0.7\textwidth}{!}{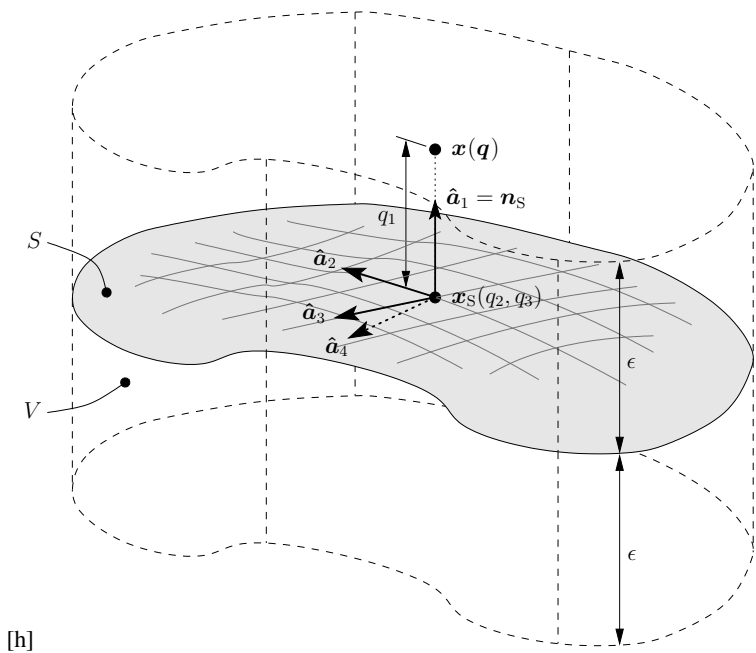}
\caption{The smooth interface surface $S$ separates two immiscible materials within a volume $V$.  Expressed in terms of the convected surface coordinates $(\scali[2]{q},\scali[3]{q})$, $S$ is constant for all time.
\label{fig:delta_function_volume}}
\end{figure}

To derive equations~(\ref{eq:area_definition}) and (\ref{eq:delta_transport_equation}) we utilise a volume $V$ that contains two immiscible materials that are separated by such a single smooth and continuous interface --- surface $S$ (see figure \ref{fig:delta_function_volume}).  The surface completely spans $V$ such that the circumference of $S$ occurs along the boundary of $V$.  As the velocity within $S$ is equal to that of the material (by equation~(\ref{eq:xsvelocity})), the boundary location of $S$ is constant for all time when expressed in terms of the convected surface coordinates $\scali[2]{q}$ and $\scali[3]{q}$.  Formally $V$ is constructed by projecting both above and below $S$ in the direction of $\vect[s]{n}$ a distance $\epsilon$.  Hence within $V$, $-\epsilon<\scali[1]{q}<\epsilon$, and provided that $\epsilon$ is small enough, the relationship between $\vect{x}$ and $\vect{q}$ expressed by equation~(\ref{eq:xtoqdef}) will be unique.

To derive equation~(\ref{eq:area_definition}) we integrate the surface delta function over $V$ and express the integral in terms of the new coordinate system $\vect{q}$:
\begin{equation}
\int_V \scal[S]{\delta}(\vect{x}) d\vect{x} = \int_V \delta(\scali[1]{q}) \left [ \frac{\partial \vect{x}}{\partial \scali[1]{q}} \cdot \left ( \frac{\partial \vect{x}}{\partial \scali[2]{q}} \cross \frac{\partial \vect{x}}{\partial \scali[3]{q}} \right ) \right ] d\vect{q}
\label{eq:deltaint1} .
\end{equation}
As the Dirac delta function $\delta(\scali[1]{q})$ is nonzero only for $\scali[1]{q}=0$, the Jacobian in this equation (the term in square brackets) need only be evaluated for $\scali[1]{q}=0$.  Hence, utilising equation~(\ref{eq:xtoqdef}) to evaluate the partial derivatives equation~(\ref{eq:deltaint1}) becomes
\begin{equation}
\int_V \scal[S]{\delta}(\vect{x}) d\vect{x} = \int_{-\epsilon}^\epsilon \delta(\scali[1]{q}) d\scali[1]{q} \int_S \vect[S]{n} \cdot \left ( \vecti[2]{a} \cross \vecti[3]{a} \right )  d\scali[2]{q}d\scali[3]{q}
\label{eq:deltaint2}
\end{equation}
where
\begin{equation}
\vecti[2]{a}=\left . \frac{\partial \vect{x}}{\partial \scali[2]{q}} \right |_{\scali[1]{q}=0}=\frac{\partial \vect[S]{x}}{\partial \scali[2]{q}} \quad \text{and} \quad \vecti[3]{a}=\left . \frac{\partial \vect{x}}{\partial \scali[3]{q}} \right |_{\scali[1]{q}=0}=\frac{\partial \vect[S]{x}}{\partial \scali[3]{q}}
\label{eq:a2a3def}
\end{equation}
are two non-conincident vectors that are tangential to $S$ at $\vect[S]{x}$.  By definition of the Dirac delta function the first integral on the right of this equation~(\ref{eq:deltaint2}) is equal to one (as $\epsilon$ is a small positive number).  For the second integral we note that the vectors $\vecti[2]{a}$ and $\vecti[3]{a}$ are both orthogonal to $\vect[S]{n}$, and hence without loss of generality we define
\begin{equation}
\vect[S]{n}=\frac{ \vecti[2]{a} \cross \vecti[3]{a} }{ \abs{ \vecti[2]{a} \cross \vecti[3]{a} } }
\label{eq:nsdef} .
\end{equation}
This allows the volume integral of $\scal[S]{\delta}$ to be written as
\begin{equation}
\int_V \scal[S]{\delta}(\vect{x}) d\vect{x} = \int_S \left | \frac{\partial \vect[S]{x}}{\partial \scali[2]{q}} \cross \frac{\partial \vect[S]{x}}{\partial \scali[3]{q}} \right |  d\scali[2]{q}d\scali[3]{q} = \int_S dS = A
\label{eq:deltaint3}
\end{equation}
where $A$ is the area of surface $S$ \citep[][p454, \S 10.6]{kreyszig06}.  Noting that any volume containing arbitrary surfaces can be composed of volumes that contain smooth and continuous surfaces and volumes that contain no surface (in which equation~(\ref{eq:deltaint3}) is trivially satisfied), equation~(\ref{eq:area_definition}) from the main text results.

\subsection{Surface delta function transport equation \label{sec:deltapropertiestransport}}

To derive equation~(\ref{eq:delta_transport_equation}) we take the derivative of equation~(\ref{eq:deltaint3}) with respect to time.  Recognising that the boundary location of $S$ is constant in terms of $\scali[2]{q}$ and $\scali[3]{q}$, the time derivative commutes into the integral giving
\begin{equation}
\frac{d}{dt} \int_V \scal[S]{\delta}(\vect{x}) d\vect{x} = \int_S \frac{d}{dt} \Bigl [ \abs{ \vecti[2]{a} \cross \vecti[3]{a} } \Bigr ] d\scali[2]{q}d\scali[3]{q} 
\label{eq:deltaint4}.
\end{equation}
Performing the differentiation and (re)introducing the Dirac delta function leads to
\begin{align}
\frac{d}{dt} \int_V \scal[S]{\delta}(\vect{x}) d\vect{x} & =  \int_{-\epsilon}^\epsilon \delta(\scali[1]{q}) d\scali[1]{q} \int_S \mathcal{A}(\vect{q}) \abs{ \vecti[2]{a} \cross \vecti[3]{a} } d\scali[2]{q}d\scali[3]{q} \label{eq:deltaint4} \\
& = \int_V \delta(\scali[1]{q}) \mathcal{A}(\vect{q}) \left [ \frac{\partial \vect{x}}{\partial \scali[1]{q}} \cdot \left ( \frac{\partial \vect{x}}{\partial \scali[2]{q}} \cross \frac{\partial \vect{x}}{\partial \scali[3]{q}} \right ) \right ] d\vect{q} \nonumber \\
& = \int_V \scal[S]{\delta}(\vect{x}) \mathcal{A}(\vect{x}) d\vect{x}
\label{eq:deltaint5}
\end{align}
where 
\begin{equation}
\mathcal{A} =  \frac{ \left ( \vecti[2]{a} \cross \vecti[3]{a} \right ) }{ \abs{ \vecti[2]{a} \cross \vecti[3]{a} }^2 } \cdot \left [ \frac{d\vecti[2]{a}}{dt} \cross \vecti[3]{a} + \vecti[2]{a} \cross \frac{d\vecti[3]{a}}{dt} \right ]
\label{eq:aaa1}
\end{equation}
is a function of the local surface geometry.

To simplify the expression for $\mathcal{A}$ we recall from equation~(\ref{eq:deltaint2}) that the surface integral in equation~(\ref{eq:deltaint4}) which contains $\mathcal{A}$ is evalulated under the condition $\scali[1]{q}=0$.  Assuming this condition we combine equations~(\ref{eq:xsvelocity}) and (\ref{eq:a2a3def}) to derive the identity,
\begin{equation}
\frac{d\vecti[2]{a}}{dt} =
\frac{d}{dt} \left ( \frac{\partial \vect[S]{x}}{\partial\scali[2]{q}} \right ) =
\frac{\partial}{\partial \scali[2]{q}} \left ( \frac{d\vect[S]{x}}{dt} \right ) =
\frac{\partial\vect{v}}{\partial\scali[2]{q}} =
\frac{\partial\vect{x}}{\partial\scali[2]{q}} \cdot \vnabla\vect{v} =
\frac{\partial\vect[S]{x}}{\partial\scali[2]{q}} \cdot \vnabla\vect{v} =
\vecti[2]{a} \cdot \vnabla\vect{v} .
\end{equation} 
Substituting this and an analygous result for ${d\vecti[3]{a}}/{dt}$ into equation~(\ref{eq:aaa1}) yields
\begin{equation}
\mathcal{A} =  \beta^2 \left ( \vecti[2]{\hat{a}} \cross \vecti[3]{\hat{a}} \right ) \cdot \bigl [ { \left ( \vecti[2]{\hat{a}}\cdot\vnabla\vect{v} \right ) \cross \vecti[3]{\hat{a}} + \vecti[2]{\hat{a}} \cross \left ( \vecti[3]{\hat{a}}\cdot\vnabla\vect{v} \right ) } \bigr ]
\label{eq:aaa2}
\end{equation}
where
\begin{equation}
\vecti[2]{\hat{a}}=\frac{\vecti[2]{a}}{\abs{\vecti[2]{a}}} \quad \text{and} \quad \vecti[3]{\hat{a}}=\frac{\vecti[3]{a}}{\abs{\vecti[3]{a}}}
\end{equation}
are unit vectors in each of the two surface coordinate directions and $\beta=\abs{\vecti[2]{\hat{a}}\cross\vecti[3]{\hat{a}}}^{-1}$.

We now define a forth unit vector $\vecti[4]{\hat{a}}$ which is coplannar with $\vecti[2]{\hat{a}}$ and $\vecti[3]{\hat{a}}$ (and hence tangential to $S$) such that the vectors $(\vect[S]{n},\vecti[2]{\hat{a}},\vecti[4]{\hat{a}})$ form a right handed coordinate system at $\vect[S]{x}$ (as shown in figure \ref{fig:delta_function_volume}).  As the orientation of $\vecti[4]{\hat{a}}$ obeys $\vect[S]{n}=\vecti[2]{\hat{a}}\cross\vecti[4]{\hat{a}}$, the new vector can be expressed as
\begin{equation}
\vecti[4]{\hat{a}}=\alpha\vecti[2]{\hat{a}}+\beta\vecti[3]{\hat{a}} , \quad \text{or} \quad
\vecti[3]{\hat{a}}=\frac{1}{\beta}\left ( \vecti[4]{\hat{a}}-\alpha\vecti[2]{\hat{a}} \right ),
\label{eq:a4def}
\end{equation}
where $\alpha$ is a finite scalar and equation~(\ref{eq:nsdef}) has been used.  Substituting equation~(\ref{eq:a4def}) into equation~(\ref{eq:aaa2}) yields after some simplification
\begin{equation}
\mathcal{A} =  \left ( \vecti[2]{\hat{a}} \cross \vecti[4]{\hat{a}} \right ) \cdot \bigl [ { \left ( \vecti[2]{\hat{a}}\cdot\vnabla\vect{v} \right ) \cross \vecti[4]{\hat{a}} + \vecti[2]{\hat{a}} \cross \left ( \vecti[4]{\hat{a}}\cdot\vnabla\vect{v} \right ) } \bigr ]
\label{eq:aaa3} .
\end{equation}
Employing the identity \citep[][p814, \S A.2]{bird02}
\begin{equation}
[\vect{u}\cross\vect{v}]\cdot[\vect{w}\cross\vect{z}]=(\vect{u}\cdot\vect{w})(\vect{v}\cdot\vect{z})-(\vect{u}\cdot\vect{z})(\vect{v}\cdot\vect{w})
\end{equation}
and noting that $\vecti[2]{\hat{a}}\cdot\vecti[4]{\hat{a}}=0$ leads to
\begin{equation}
\mathcal{A} = \vecti[2]{\hat{a}}\vecti[2]{\hat{a}}:\vnabla\vect{v}+ \vecti[4]{\hat{a}}\vecti[4]{\hat{a}}:\vnabla\vect{v} = (\tens{I}-\vect[S]{n}\vect[S]{n}):\vnabla\vect{v}
\label{eq:aaa4}
\end{equation}
where $\tens{I}$ is the unit tensor.

With $\mathcal{A}$ defined we return to the development of equation~(\ref{eq:deltaint5}):  Noting that for $\scal[S]{\delta}\ne0$ the boundary of $V$ moves at the local material velocity $\vect{v}$, the left hand side of this equation can be expanded using the Leibnitz formula for differentiating a volume integral and the Gauss-Ostrogradskii divergence theorem.  On the right hand side we substitute $\mathcal{A}$ from equation~(\ref{eq:aaa4}).  These operations yield
\begin{equation}
\int_V \frac{\partial \scal[S]{\delta}}{\partial t} d\vect{x} + \int_V \vnabla \cdot \left ( \scal[S]{\delta}\vect{v} \right ) d\vect{x} = \int_V \scal[S]{\delta} (\tens{I}-\vect[S]{n}\vect[S]{n}):\vnabla\vect{v} d\vect{x}
\label{eq:deltaint6} .
\end{equation}
Equation~(\ref{eq:deltaint6}) is valid for a specific volume geometry that contains a single smooth and continuous interface surface.  However, equation~(\ref{eq:deltaint6}), like equation~(\ref{eq:deltaint3}), is trivially satisfied in volumes that contain no interface surface.  As any volume can be composed of volumes that contain a smooth and continuous interface surface and volumes that contain no interface surface, equation~(\ref{eq:deltaint6}) must hold for any arbitrary volume.  Hence, equation~(\ref{eq:deltaint6}) must hold at all locations and equation~(\ref{eq:delta_transport_equation}) from the main text results.

\Citet{james04}, based on the work of \citet[][p132]{batchelor67}, give an equation that is similar to equation~(\ref{eq:delta_transport_equation}) but expressed using non-equivalent terms: The equation of \citet{james04} concerns the material derivative of an infinitesimal surface element, whereas equation~(\ref{eq:delta_transport_equation}) concerns the material derivative of the surface delta function.

\subsection{Surface integral of surface delta function \label{sec:deltapropertiessurface}}

In order to apply the mechanical energy balance of equation (\ref{eq:macroscopic_balance}) we also need to be able to evaluate the integral of the surface delta function over a surface.  In the following we derive equation (\ref{eq:line_definition}) that is given in the main text for this purpose.

\begin{figure} 
\subfloat[\label{fig:2Ddeltaiso}]{\def\svgwidth{0.6\textwidth}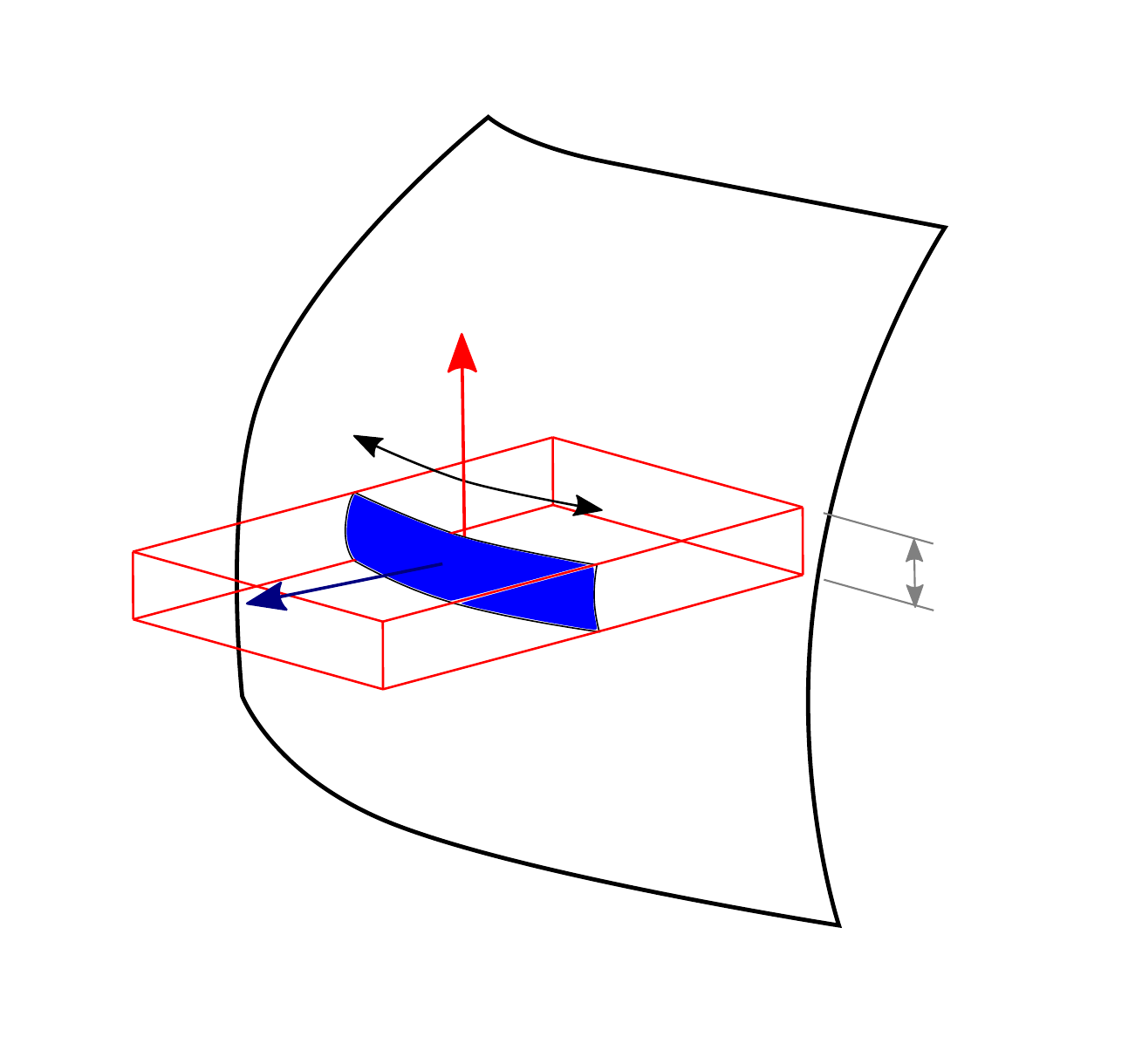}%
\subfloat[\label{fig:2Ddeltatriangle}]{\def\svgwidth{0.4\textwidth}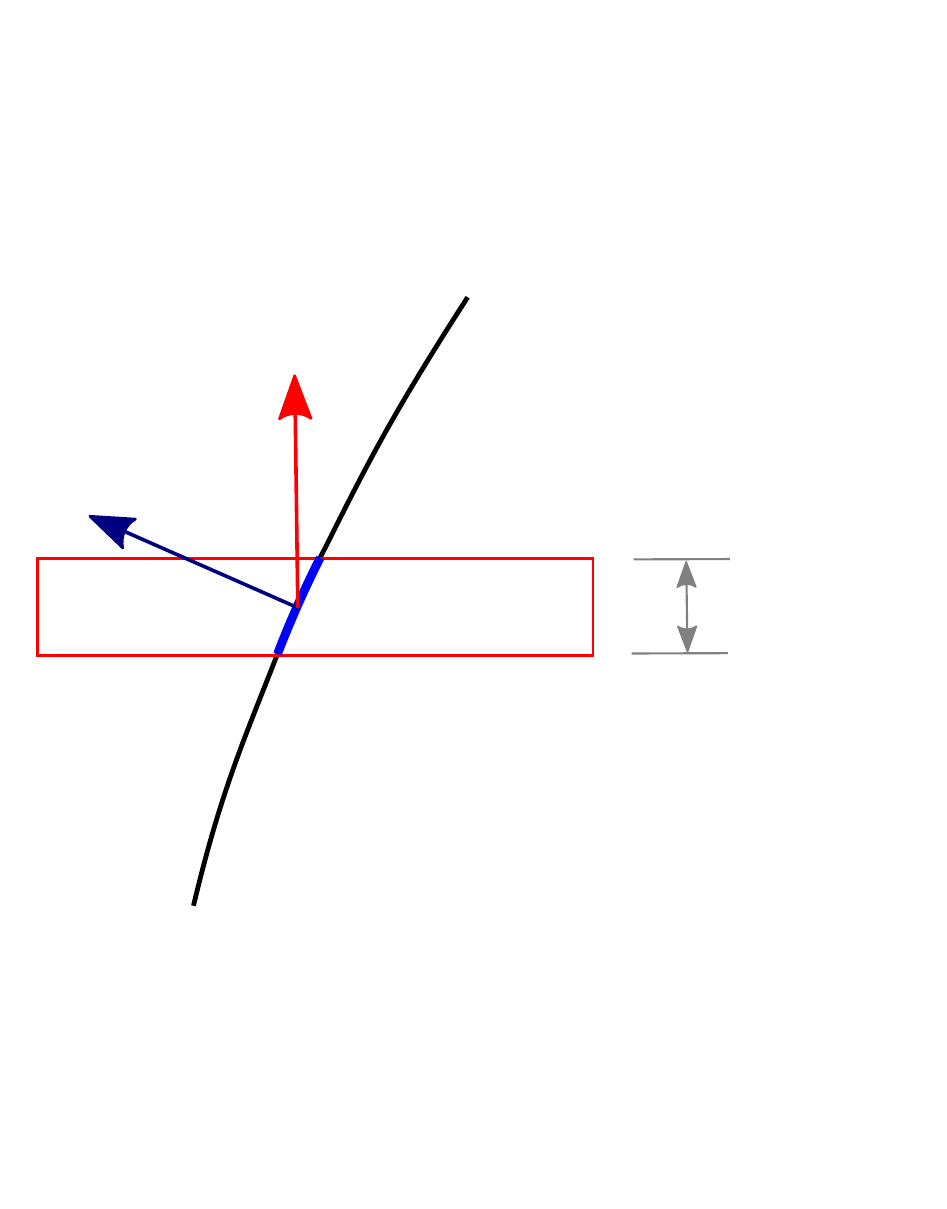}%
\caption{The two-dimensional integral of the delta function \label{fig:2Ddelta}}
\end{figure}

The derivation is based on a small amount of interfacial area, $\Delta A$, that has a length of $\Delta l$, and that is contained within a small element of volume $\Delta V$.  The volume element is thin in the direction of $\vect{n}$, having a thickness of $\epsilon$ in this direction, so that $\Delta V=\epsilon \Delta S$.  $\Delta S$ is a small part of a larger surface, $S$. Figures \ref{fig:2Ddeltaiso} and \ref{fig:2Ddeltatriangle} describe this geometrical system using a projection and cross-section, respectively.

The starting point for the analysis is the definition of the surface delta function, that is equation (\ref{eq:area_definition}), applied to the small volume $\Delta V$.  This gives
\begin{equation}
\Delta A = \int_{\Delta V} \scal[S]{\delta} dV = \int_{-\epsilon/2}^{\epsilon/2} \int_{\Delta S} \scal[S]{\delta} dS d \scali[\vect{n}]{x} \approx \epsilon \int_{\Delta S} \scal[S]{\delta} dS
\label{eq:surfdelta1}
\end{equation}
where $\scali[\vect{n}]{x}$ is a coordinate in the direction of $\vect{n}$ centred on $\Delta A$.  The last equality holds true for small $\epsilon$.  Meanwhile for small $\Delta A$ $\vect[S]{n}$ is approximately uniform over $\Delta A$ and the geometry of the intersecting surfaces gives (see Figure \ref{fig:2Ddeltatriangle})
\begin{equation}
\Delta A = \frac{\Delta l \epsilon}{\sqrt{1-(\vect[S]{n} \cdot \vect{n})^2}}
\label{eq:surfdelta2}
\end{equation}
Equating equations (\ref{eq:surfdelta1}) and (\ref{eq:surfdelta2}) leads to
\begin{equation}
\int_{\Delta S} \scal[S]{\delta} dS = \frac{\Delta l}{\sqrt{1-(\vect[S]{n} \cdot \vect{n})^2}}
\label{eq:surfdelta3}
\end{equation}
Finally, recognising that $\Delta S$ is a small section of a larger surface $S$, and that $\Delta l$ is a small section of the entire intersection between $S$ and $\scal[S]{\delta}$ that has a length of $l$, equation (\ref{eq:surfdelta3}) can be generalised in the limit of $\Delta l \rightarrow 0$ to
\begin{equation}
\int_{S} \scal[S]{\delta} dS = \int_0^l \frac{dl'}{\sqrt{1-(\vect[S]{n} \cdot \vect{n})^2}}
\label{eq:surfdelta4}
\end{equation}
where $l'$ is a pathlength parameterisation of the curve that is defined as the intersection between $S$ and $\scal[S]{\delta}$, and $\vect[S]{n} \cdot \vect{n}$ is a function of $l'$.  Equation (\ref{eq:surfdelta4}) is the most general form of this surface delta identity, however by assuming that $\vect[S]{n} \cdot \vect{n}$ is independent of the pathlength $l'$ equation (\ref{eq:line_definition}) from the main text results.

\section{Appendix: Dissipation event order of magnitude energy analysis \label{sec:dissipationordermag}}

As discussed in the main text, the objective of this section is to quantify in an order of magnitude sense what energy terms are significant during the dissipation events so that we can determine the ultimate destination for the specific surface energy $\dsDeltasigma$ that is liberated from the dissipation events.

Summing the individual dissipation event energy balances expressed by equation (\ref{eq:balhat2}) over all $N$ dissipation events and substituting the order of magnitude of $\sum_{k=1}^N \dsT_{0,k}$ from equation (\ref{eq:dsT04}) gives
\begin{equation}
\order{\rho \vcv^2 \hrough} + \order{\rho g \hrough^2} + \dsDeltasigma = \sum_{k=1}^N \dsT_{0,k} = \sum_{i=1}^6 \sum_{k=1}^N \dsT_{i,k}
\label{eq:dsbal1}
\end{equation}
We now examine each of the six $\sum_{k=1}^N \dsT_{i,k}$ terms on the RHS of this equation.

For $\dsT_{1,k}$, as for the equilibrium analysis $\ncv \cdot \vvcv$ on $\Scvend$, so the only contribution to this term comes from $\Scvcir$.  However on the circumference of the CV the interface is not affected by the dissipation event (as it is $\order{\rcv}$ away from $\xcap$) so that the value of the surface integral is independent of whether the time is within an equilibrium or dissipation stage.  Noting then that $\sum_{k=1}^{N} \int_{\dsDtk} = \taucap$ and $\sum_{k=1}^{N+1} \int_{\eqDtk} = \tau-\taucap$ the first dissipation term can be written as
\begin{align}
\sum_{k=1}^{N} \dsT_{1,k} & = \frac{1}{\Acv} \sum_{k=1}^{N} \int_{\dsDtk} \int_{\Scv} \sum_{i<j} \sigmaij \scali[\text{S},ij]{\delta} \ncv \cdot \vvcv dS dt \nonumber \\
& = \frac{\taucap}{\tau-\taucap} \sum_{k=1}^{N+1} \eqT_{1,k} = \order{\frac{\taucap}{\tau} \sum_{k=1}^{N+1} \eqT_{1,k}} \label{eq:dsT1}
\end{align}
where equation (\ref{eq:taucapdtau}) has been utilised.

The second term on the RHS of equation (\ref{eq:dsbal1}) is concerned with movement of kinetic energy through the CV boundary during the dissipation periods.  Using the stationary solid model, this term is
\begin{equation}
\sum_{k=1}^{N} \dsT_{2,k} = \frac{1}{\Acv} \sum_{k=1}^{N} \int_{\dsDtk} \int_{\Scvfluid} \ncv \cdot \frac{1}{2}\rho {|\vdsv|}^2 ( \vvcv - \vdsv ) dS dt \label{eq:dsT21}
\end{equation}
By recognising the separation of velocities $\vcv \ll \vcap$ when using equation (\ref{eq:dsv}) in the above, for order of magnitude purposes products of $\vcv$ and $\vcap$ can be ignored and the above term written as a contribution from the continuous and dissipation interface movements as
\begin{align}
\sum_{k=1}^{N} \dsT_{2,k} & = \order{\frac{\taucap}{\tau-\taucap} \sum_{k=1}^{N+1} \eqT_{2,k}} + \order{\frac{\rho \vcap^3 \taucap \hrough^3}{\Acv \lcv}} \nonumber \\
& = \order{\frac{\taucap}{\tau}\sum_{k=1}^{N+1} \eqT_{2,k}} + \order{\frac{\rho \vcap^2 \hrough^2}{\lcv}}
\label{eq:dsT22}
\end{align}
In evaluating the dissipation event term on the first line of this equation (the second on RHS), consistent with the dissipation fluid model of equation (\ref{eq:dsv}) $\vdsv$ is only significant within an area of $\order{\hrough^2}$ on each end of the CV, and as the dissipation events are evenly distributed over the length $\lcv$ of the CV, only a proportion $\order{\hrough / \lcv}$ of the dissipation events contribute to this surface integral.  Further, from equation (\ref{eq:taucapdtau}) we have used $\taucap \vcap = \order{\lcv \Xcv / \hrough}$.

The third term on the RHS of equation (\ref{eq:dsbal1}) represents transport of gravitational potential energy over the CV boundary and is evaluated in a very similar fashion to the second term.  Recognising as previously that energy transport due to specific dissipation movements only occurs over the central $\order{\hrough^2}$ area of $\Scvend$, and that during these events $\hat{\Phi}=\order{g \hrough}$, this term can be evaluated as
\begin{align}
\sum_{k=1}^{N} \dsT_{3,k} & = \frac{1}{\Acv} \sum_{k=1}^{N} \int_{\dsDtk} \int_{\Scvfluid} \ncv \cdot \rho \hat{\Phi} ( \vvcv - \vdsv )  dS dt \nonumber \\
& = \order{\frac{\taucap}{\tau}\sum_{k=1}^{N+1} \eqT_{3,k}} + \order{\frac{\rho g \hrough^3}{\lcv}}
\label{eq:dsT31}
\end{align}
where assumptions consistent with those used for equation (\ref{eq:dsT22}) have been employed, and the first and second terms on the RHS of the second line of this equation represent contributions from the continuous and dissipation velocities occurring during the dissipation periods, respectively.

Term $\sum_{k=1}^{N} \dsT_{4,k}$, like $\sum_{k=1}^{N} \dsT_{1,k}$, is concerned with interfacial behaviour at the CV boundary.  Recognising that only the interface between the two fluids can experience a non-zero velocity, and that dissipation interfacial movements are again confined to an area of $\order{\hrough^2}$ on $\Scvend$ this term can be written as
\begin{align}
\sum_{k=1}^{N} \dsT_{4,k} & =- \frac{1}{\Acv} \sum_{k=1}^{N} \int_{\dsDtk} \int_{\Scvfluid} \scali[12]{\sigma} \scali[\text{S,}12]{\delta} \vecti[\text{S,}12]{n} \vecti[\text{S,}12]{n} : \vdsv \ncv  dS dt \nonumber \\
& = \order{\frac{\taucap}{\tau}\sum_{k=1}^{N+1} \eqT_{4,k}} + \order{\sigma \frac{\hrough}{\lcv}}
\label{eq:dsT41}
\end{align}
where $\taucap \vcap$ has again been evaluated using equation (\ref{eq:taucapdtau}).

The final two terms from the RHS of equation (\ref{eq:dsbal1}) are concerned with material stresses, specifically representing the boundary work and internal viscous dissipation occurring on and within the CV, respectively.  As per previously in the absence of solid velocities these terms only have contributions from the fluid regions.  For the boundary work term we substitute the Newtonian stress equation (\ref{eq:newtonian_stress}) and dissipation period velocity, velocity gradient and pressure expressions of equations (\ref{eq:dsv}), (\ref{eq:dsnablav}) and (\ref{eq:dsp}), respectively, into equation (\ref{eq:T5}) applied over the dissipation periods, giving
\begin{align}
\sum_{k=1}^{N} \dsT_{5,k} & = - \frac{1}{\Acv} \sum_{k=1}^{N} \int_{\dsDtk} \int_{\Scvfluid} \dsp \vdsv \cdot \ncv dS dt \nonumber \\
& + \frac{1}{\Acv} \sum_{k=1}^{N} \int_{\dsDtk} \int_{\Scvfluid} \mu \left [ \dsnablav + (\dsnablav)^T \right ] : \vdsv \ncv dS dt \nonumber \\
& = \order{\frac{\taucap}{\tau-\taucap} \sum_{k=1}^{N+1} \eqT_{5,k}} + \mathcal{O}\left ( \frac{\vcap \taucap \hrough}{\Acv \lcv} \right . \nonumber \\
& \left . \int_{\Scvend,r<\hrough} \left [ \frac{\mu \vcap}{\max(r,\hmol)} + \rho \vcap^2 + \frac{\sigma}{\hrough} \right ] dS \right )
\label{eq:dsT51}
\end{align}
The final integral in this equation is taken over the ends of the CV that are within $\order{\hrough}$ of its centreline, with $r$ representing the distance to the centreline.  Evaluating this integral leads to
\begin{align}
\sum_{k=1}^{N} \dsT_{5,k} & = \order{ \frac{\taucap}{\tau} \sum_{k=1}^{N+1} \eqT_{5,k} } \nonumber \\
& + \order{ \rho \vcap^2 \frac{\hrough^2}{\lcv}} + \order{ \mu \vcap \frac{\hrough}{\lcv}} + \order{\sigma\frac{\hrough}{\lcv}}
\label{eq:dsT52}
\end{align}
The viscous dissipation term $\sum_{k=1}^{N} \dsT_{6,k}$ is evaluated similarly.  Noting that velocities within the solid are zero, and that the fluid is incompressible, as for the corresponding equilibrium term analysis there is no contribution from the pressure term $\dsp$ and the dissipation velocity gradient can be integrated over a region of volume $\order{\hrough^3}$ within the fluid, giving
\begin{align}
\sum_{k=1}^{N} \dsT_{6,k} & = - \frac{1}{\Acv} \sum_{k=1}^{N} \int_{\dsDtk} \int_{\Vcvfluid} \mu \left [ \dsnablav + (\dsnablav)^T \right ] : \dsnablav dV dt \nonumber \\
& = \order{ \frac{\taucap}{\tau} \sum_{k=1}^{N+1} \eqT_{6,k} } + \orders{ \mu \vcap \ln \left ( \frac{\hrough}{\hmol} \right ) }
\label{eq:dsT61}
\end{align}

Equation (\ref{eq:dsbal2}) in the main text results from substituting equations (\ref{eq:dsT1}), (\ref{eq:dsT22}), (\ref{eq:dsT31}), (\ref{eq:dsT41}), (\ref{eq:dsT52}) and (\ref{eq:dsT61}) into equation (\ref{eq:dsbal1}).

\bibliography{abbreviated,droplet,daltonh,suspend,microfluidics,main}
\end{document}